\documentclass[a4paper,10pt]{article}
\usepackage{latexsym}
\usepackage{amsmath}
\usepackage{amssymb}
\usepackage{amscd}

\newlength{\minitwocolumn}
\makeatletter
\@addtoreset{equation}{section}
\makeatother

\title{Difference equations for 
correlation functions of \\
$A^{(1)}_{n-1}$-face model
with boundary reflection}

\author{Yas-Hiro Quano\thanks
{email: quanoy@suzuka-u.ac.jp}}

\date{\it Department of Medical Electronics, 
Suzuka University of Medical Science \\
      \it Kishioka-cho, Suzuka 510-0293, Japan}
\begin{document}

\maketitle
\begin{abstract}
The $A^{(1)}_{n-1}$-face model with boundary reflection is 
considered on the basis of the boundary CTM bootstrap. 
We construct the fused boundary Boltzmann weights 
to determine the normalization factor. 
We derive difference equations of  
the quantum Knizhnik-Zamolodchikov type for 
correlation functions of the boundary model. 
The simplest difference equations 
are solved in the case of the free boundary condition. 
\end{abstract}

\section{Introduction}

Many two dimensional integrable models without boundary 
have been solved by using the representation theory 
of affine quantum group \cite{JM,FJMMN}. 
The integrability of bulk models 
is ensured by the factorized scattering condition 
or the Yang--Baxter equation, in addition to the 
unitarity and crossing symmetry condition \cite{ZZ,ESM}. 

Cherednik \cite{RE,bKZ} showed that the integrability 
in the presence of reflecting boundary is ensured by 
the reflection equation (boundary Yang--Baxter 
equation) and the Yang--Baxter equation for bulk theory. 
A systematic treatment of determining the spectrum of 
integrable models with boundary reflection was 
initiated by Sklyanin \cite{Skl} in the framework 
of the algebraic Bethe Ansatz. 
The boundary interaction is specified by the boundary 
$S$-matrix for massive quantum theories \cite{GZ}, 
by the reflection matrix 
for lattice vertex models \cite{Skl}, and 
by the boundary weights for lattice face models 
\cite{Kul,BPO}. 

In our previous paper \cite{bBela} 
Belavin's $\mathbb{Z}_n$-symmetric elliptic vertex 
model with boundary reflection is considered on the 
basis of the boundary CTM(corner transfer matrix) bootstrap 
formulated in \cite{JKKMW}. We derived a set of difference 
equations called the boundary quantum Knizhnik-Zamolodchikov 
equation for correlation functions in the boundary Belavin's 
model. Furthermore, we obtained the boundary spontaneous 
polarization by solving the simplest difference equations. 
The resulting quantities are exactly equal to the square 
of that for bulk spontaneous polarization \cite{SPn}, 
up to a phase factor. 

In this paper we consider the $A^{(1)}_{n-1}$-face model 
\cite{JMO} with a boundary on the basis of boundary 
CTM bootstrap. The $\mathbb{Z}_n$-symmetric model 
and the $A^{(1)}_{n-1}$-face model are related by the 
vertex-face correspondence \cite{JMO} in the bulk theory. 
Thus we wish to find the similar structure in the 
boundary $A^{(1)}_{n-1}$-face model as the one 
observed in \cite{bBela}. 

Integrable face models on a half infinite lattice 
have been studied in \cite{MW,BFKZ,BFKSZ,H}. 
In \cite{MW} the transfer matrix of the boundary 
ABF model (the boundary $A^{(1)}_{1}$-face model) 
was diagonalized by constructing the boundary vacuum 
states, under the Ansatz \cite{JKKKM} that the boundary 
vacuum states should be obtained from the Fock vacuums 
by applying the exponential of the infinite sum of the 
quadratic bosonic oscillators associated with 
the bulk ABF models \cite{LuP}. 
In \cite{BFKZ} the solution to the reflection equation 
was given for the boundary $A^{(1)}_{n-1}$-face model. 
In \cite{BFKSZ} free energy and critical 
exponents was obtained for $n=2$, the boundary 
ABF model case. In order to discuss the higher $n$ cases 
we shall establish the boundary crossing symmetry in 
section 3. In \cite{H} correlation functions for the 
boundary $XYZ$ model were obtained in terms of those 
for the boundary ABF model \cite{MW} by using the 
vertex-face transformation method \cite{LaP}. 

The rest of this paper is organized as follows.
In section 2 we review the boundary $A^{(1)}_{n-1}$-face 
model, thereby fixing our notation. In section 3 we 
introduce the fusions of $K$-matrix to determine the 
normalization factors of $K$-matrix. We also 
establish the boundary crossing symmetry. 
In section 4 we construct lattice realization of the 
vertex operators on the basis of the 
boundary CTM bootstrap approach. Furthermore, we derive 
difference equations for correlation functions of 
the boundary $A^{(1)}_{n-1}$-face model. 
In section 5 we give some concluding remarks. 
In Appendix A we give the explicit expressions of 
fused Boltzmann weights of the bulk $A^{(1)}_{n-1}$-face 
model \cite{AJMP}. In Appendix B we give a simple sketch 
of the proof of the reflection equation. 

\section{Boundary $A^{(1)}_{n-1}$-face model}

The present section aims to formulate 
the problem, thereby fixing the notation. 

\subsection{Theta functions} 

Throughout this paper we fix the integers $n$ and 
$r$ such that $r\geqq n+2$, and also fix 
the parameter $x$ such that $0<x<1$. 
We will use the abbreviations, 
\begin{equation}
[z]=x^{\frac{z^2}{r}-z}\Theta_{x^{2r}}(x^{2z}), 
\end{equation}
where the Jacobi theta function is given by 
\begin{eqnarray}
\Theta_{q}(\zeta )&=&(\zeta ; q)_\infty 
(q\zeta^{-1}; q)_\infty (q, q)_\infty , \\
(\zeta ; q_1 , \cdots , q_m )&=& 
\prod_{i_1 , \cdots , i_m \geqq 0} 
(1-\zeta q_1^{i_1} \cdots q_m^{i_m}). 
\end{eqnarray}
For an additive parameter $z$, we often use the 
corresponding multiplicative parameter $\zeta=x^{2z}$. 

For later conveniences we also 
introduce the following symbols
\begin{eqnarray}
r_{m}(z)&=&\zeta^{-\frac{r-1}{r}\frac{n-m}{n}}
\frac{g_{m}(\zeta )}{g_{m}(\zeta^{-1})}, ~~~~
g_{m}(\zeta )=
\frac{\{x^{2n+2r-m-1}\zeta\}
\{x^{m+1}\zeta\}}
{\{x^{2n-m+1}\zeta\}\{x^{2r+m-1}\zeta\}}, 
\end{eqnarray}
where $1\leqq m\leqq n-1$ and 
\begin{equation}
\{\zeta \}=(\zeta ;x^{2n},x^{2r})_\infty . 
\label{eq:{z}}
\end{equation}
In particular the function $r_1 (z)$ will appear 
in the expression of the Boltzmann weights of the 
$A^{(1)}_{n-1}$-face model in the regime I\!I\!I. 

\subsection{The weight lattice of $A^{(1)}_{n-1}$}

Let $V=\mathbb{C}^n$ and 
$\{ \varepsilon _\mu \}_{0 \leqq \mu \leqq n-1}$ be 
the standard orthogonal basis with the inner 
product $\langle \varepsilon _\mu , 
\varepsilon _\nu \rangle =\delta_{\mu \nu}$. 
The weight lattice of $A^{(1)}_{n-1}$ 
is defined as follows: 
\begin{equation}
P=\bigoplus_{\mu =0}^{n-1} 
\mathbb{Z} \bar{\varepsilon}_\mu , 
\label{eq:wt-lattice}
\end{equation}
where 
$$
\bar{\varepsilon}_\mu =
\varepsilon _\mu -\varepsilon , 
~~~~\varepsilon =\frac{1}{n}\sum_{\mu =0}^{n-1} 
\varepsilon _\mu . 
$$
We denote the fundamental weights 
by $\omega_\mu\,(1\leqq \mu \leqq n-1)$ 
$$
\omega_\mu =\bar{\varepsilon }_0 + 
\bar{\varepsilon }_1 + \cdots \bar{\varepsilon }_{\mu -1}. 
$$
Since $\omega_n =0$, you can define $\omega_\mu$ 
for $\mu\in\mathbb{Z}$ by setting $\omega_{\mu +n}=
\omega_\mu$. 
For $a\in P$ we set 
\begin{equation}
a_{\mu\nu}=\bar{a}_\mu -\bar{a}_\nu , ~~~~
\bar{a}_\mu =\langle a+\rho , 
\varepsilon _\mu \rangle , ~~~~ 
\rho =\sum_{\mu =1}^{n-1} \omega_\mu . 
\end{equation}
We also set 
$$
P^+_l =\left\{ \left. 
\sum_{\mu =1}^{n-1} a^\mu \omega_\mu \right| 
a^1 , \cdots , a^{n-1}\in\mathbb{Z}_{\geqq 0}, 
\sum_{\mu =1}^{n-1} a^\mu \leqq l \right\}. 
$$
We may denote $a\in P^+_l$ by 
$a=\displaystyle\sum_{\mu =1}^{n-1} 
a^\mu \omega_\mu +\left( l-\sum_{\mu =1}^{n-1} 
a^\mu \right) \omega_0$. 

\subsection{The $A^{(1)}_{n-1}$ face model}

The $A^{(1)}_{n-1}$ face model is the one 
whose local state $a$ is restricted such that 
$a\in P^+_{r-n}$. 
An ordered pair $(a,b) \in P^2$ 
is called {\it admissible} if $b=a+\bar{\varepsilon}_\mu$, 
for a certain $\mu\,(0\leqq \mu \leqq n-1)$. 
In what follows we denote $b\longleftarrow a$ when 
$(a,b) \in P^2$ is admissible. Furthermore, an 
ordered quartet $(a,b,c,d) \in P^4$ is called 
{\it admissible} if the four pairs $(a,b), (a,d), 
(b,c)$ and $(d,c)$ are admissible. 

For $(a, b, c, d)\in P^4$ let 

\unitlength 1mm
\begin{picture}(100,20)
\put(20,7){$W(a,b,c,d|z_1 -z_2 )=W
\left( \left. \begin{array}{cc} c & d 
\\ b & a \end{array} 
\right| z_1 -z_2 \right)=$} 
\put(100,3){\begin{picture}(101,0)
\put(10,0){\vector(-1,0){10}}
\put(10,0){\vector(0,1){10}}
\put(0,0){\vector(0,1){10}}
\put(10,10){\vector(-1,0){10}}
\multiput(-3,5)(2.2,0){7}{\line(1,0){1.2}}
\put(12.,5){\vector(1,0){1}}
\put(13.8,4.2){$z_1$}
\multiput(5,-3)(0,2.2){7}{\line(0,1){1.2}}
\put(5,12){\vector(0,1){1}}
\put(4,14){$z_2$}
\put(-2,10.5){$c$}
\put(10.5,10.1){$d$}
\put(10.5,-1.5){$a$}
\put(-2.3,-1.8){$b$}
\end{picture}
}
\end{picture}

\noindent be the Boltzmann weight of the 
$A^{(1)}_{n-1}$ model for the state configuration 
$(a,b,c,d)$ round a face. 
Here the four states $a, b, c$ and $d$ are 
ordered clockwise from the SE corner, and the oriented 
broken lines in the above figure carry spectral 
parameters. In this model $W(a,b,c,d|z) =0$ 
unless the quartet $(a,b,c,d)$ is admissible. 
Non-zero Boltzmann weights are parametrized in terms of 
the elliptic theta function of the spectral parameter $z$ 
as follows: 
\begin{equation}
\begin{array}{rcl}
W
\left( \left. \begin{array}{cc} 
a + 2 \bar{\varepsilon }_\mu & a+\bar{\varepsilon }_\mu \\ 
a+\bar{\varepsilon }_\mu & a \end{array} \right| 
z \right) 
& = & r_1 (z), \\[8mm]
W
\left( \left. \begin{array}{cc} 
a+\bar{\varepsilon }_\mu +\bar{\varepsilon }_\nu 
& a+\bar{\varepsilon }_\mu \\ 
a+\bar{\varepsilon }_\nu & a 
\end{array} \right| z \right) & = & r_1 (z)
\dfrac{[z][a_{\mu\nu}-1]}{[z+1][a_{\mu\nu}]} 
~~~~(\mu \neq \nu ), \label{eq:BW} \\[8mm]
W
\left( \left. \begin{array}{cc} 
a+\bar{\varepsilon }_\mu +\bar{\varepsilon }_\nu 
& a+\bar{\varepsilon }_\nu \\ 
a+\bar{\varepsilon }_\nu & a 
\end{array} \right| z \right) & = & r_1 (z)
\dfrac{[z+a_{\mu\nu}][1]}{[z+1][a_{\mu\nu}]} 
~~~~(\mu\neq \nu), 
\end{array}
\end{equation}
where $-\frac{n}{2}<z<0$ in the regime I\!I\!I. 

The Boltzmann weights (\ref{eq:BW}) solve the face-type 
Yang-Baxter equation \cite{JMO}: 
\begin{equation}
\begin{array}{cc}
~ & 
\displaystyle \sum_{g} 
W\left( \left. 
\begin{array}{cc} d & e \\ c & g \end{array} \right| 
z_1 \right)
W\left( \left. 
\begin{array}{cc} c & g \\ b & a \end{array} \right| 
z_2 \right)
W\left( \left. 
\begin{array}{cc} e & f \\ g & a \end{array} \right| 
z_1 -z_2 \right) \\[8mm]
= & \displaystyle \sum_{g} 
W\left( \left. 
\begin{array}{cc} g & f \\ b & a \end{array} \right| 
z_1 \right)
W\left( \left. 
\begin{array}{cc} d & e \\ g & f \end{array} \right| 
z_2 \right)
W\left( \left. 
\begin{array}{cc} d & g \\ c & b \end{array} \right| 
z_1 -z_2 \right). 
\end{array} \label{eq:WYBE}
\end{equation}

Some numerical calculations concerning the hard hexagon 
model in \cite{ESM} suggest that the corner transfer 
matrix (CTM) is well defined in the thermodynamic limit 
if the normalization factor $r_1 (z)$ is chosen such that 
the partition function per site is equal to unity. 
In order to fix $r_1 (z)$ the following two inversion 
relations are useful \cite{JMO}: 
\begin{equation}
\sum_g W\left( \left. 
\begin{array}{cc} c & g \\ b & a \end{array} \right| 
-z \right)
W\left( \left. 
\begin{array}{cc} c & d \\ g & a \end{array} \right| 
z \right) =\delta_{bd}, 
\label{eq:1st}
\end{equation}
\begin{equation}
\sum_g G_g W\left( \left. 
\begin{array}{cc} g & b \\ d & c \end{array} \right| 
-n-z \right)
W\left( \left. 
\begin{array}{cc} g & d \\ b & a \end{array} \right| 
z \right) =\delta_{ac}\frac{G_b G_d}{G_a}, 
\label{eq:2nd}
\end{equation}
where 
$$
G_a =\prod_{0\leqq \mu < \nu \leqq n-1} 
[a_{\mu\nu}]. 
$$
{}From the inversion trick based on these relations 
we get the expression of $r_1 (z)$ in the regime I\!I\!I. 

The Boltzmann weights (\ref{eq:BW}) also have 
$\sigma$-invariance \cite{JMO}: 
\begin{equation}
W\left( \left. 
\begin{array}{cc} 
\sigma (c) & \sigma (d) \\ 
\sigma (b) & \sigma (a) \end{array} \right| 
z \right)=W\left( \left. 
\begin{array}{cc} c & d \\ b & a \end{array} \right| 
z \right)
\label{eq:s-inv}
\end{equation}
where $\sigma$ is the diagram automorphism of 
$A^{(1)}_{n-1}$ defined by $\sigma (\omega_\mu )
=\omega_{\mu +1}$. 

\subsection{Solution to the reflection equation}

Let us consider the interation at the boundary, 
which is specified by the boundary Boltzmann weight or 
the $K$-matrix: 

\unitlength 1mm
\begin{picture}(100,20)
\put(32,7){$K(a,b,c|z)=\,
K\left( \left. a \begin{array}{c} b \\ c 
\end{array} \right| z \right)
\,=$} 
\put(80,3){\begin{picture}(101,0)
\put(12,-2){\vector(-1,1){7}}
\put(12,12){\vector(-1,-1){7}}
\put(12,-2){\line(0,1){14}}
\multiput(4,-3)(1.0,1.0){8}{\circle*{0.5}}
\put(11,4){\vector(1,1){1}}
\multiput(12,5)(-1.0,1.0){8}{\circle*{0.5}}
\put(5,12){\vector(-1,1){1}}
\put(2,4){$a$}
\put(13,11){$b$}
\put(13,-3){$c$}
\end{picture}
}
\end{picture}

\noindent Here $(b,a)$ and $(c,a)$ are admissible. 
The integrability condition of the boundary face model 
is the face-type reflection equation \cite{Kul,BPO}: 

\begin{equation}
\begin{array}{cl}
&\displaystyle\sum_{d,e} 
K\left( \left. f \begin{array}{c} g \\ e 
\end{array} \right| z_2 \right)
W\left( \left. \! \begin{array}{cc} 
c & f \\ d & e \end{array} \right| z_1 +z_2 \right)
K\left( \left. d \begin{array}{c} e \\ a 
\end{array} \right| z_1 \right) 
W\left( \left. \! \begin{array}{cc} 
c & d \\ b & a \end{array} \right| z_1 -z_2 \right) 
\\[7mm]
=& \displaystyle\sum_{d,e} 
W\left( \left. \! \begin{array}{cc} 
c & f \\ d & g \end{array} \right| z_1 -z_2 \right)
K\left( \left. d \begin{array}{c} g \\ e 
\end{array} \right| z_1 \right)
W\left( \left. \! \begin{array}{cc} 
c & d \\ b & e \end{array} \right| z_1 +z_2 \right)
K\left( \left. b \begin{array}{c} e \\ a 
\end{array} \right| z_2 \right). 
\end{array}
\label{eq:faceRE}
\end{equation}

For the bulk Boltzmann weights of the 
$A^{(1)}_{n-1}$-face model \cite{JMO} the diagonal 
solution to eq. (\ref{eq:faceRE}) is given as 
follows \cite{BFKZ}: 
\begin{equation}
K\left( \left. a+\bar{\varepsilon}_\mu 
\begin{array}{c} a \\ a' 
\end{array} \right| z \right)=f_a (z) 
\dfrac{[\bar{a}_\mu+\eta +z]}{[\bar{a}_\mu+\eta -z]} 
\delta_{aa'}, 
\label{eq:K-sol}
\end{equation}
where $\eta=\eta (a)$ may depend on $a$ but a constant 
with respect to $z$. In the present paper we take 
an $a$-independent constant $\eta$ for simplicity. 
The paper \cite{MW} employed the diagonal $K$-matrix 
for the boundary ABF model ($n=2$ case) with 
$\eta (a)=k/2 +c$ for $a=(k-1)\omega_1 +(r-1-k)\omega_0$. 
We also notice that the opposite admissible conditions 
such that $(a,b)$ and $(a,c)$ are admissible were used 
in \cite{BFKZ}. Thus we give a simple sketch of the proof 
of the claim that (\ref{eq:K-sol}) solves the reflection 
equation (\ref{eq:faceRE}) in Appendix B. 

Let $a\in P^+_{r-n}$ be the local state. 
Then in regime I\!I\!I any ground state configuration 
is labeled by some $b\in P^+_{r-n-1}$, 
where the cyclic sequence of 
$b, b+\omega_1 , \cdots , b+\omega_{n-1}$ \cite{JMO}. 
In the `low temperature' limit $x\rightarrow 0$, 
one of such ground states is realized. 
In what follows we fix one of them (say, labeled 
by $b$) and define all the correlation functions 
in terms of the `low-temperature' series expansion, 
the formal power series with respect to $x$.   
Then the fixed ground state configuration 
gives the lowest order. 
Furthermore, any finite order contribution 
results from the configurations which differ from 
that of the fixed ground state 
by altering a finite number of local states. 
Thus the infinite number of states at far enough 
sites should coincide with the fixed ground state 
configuration labeled by $b$. Such one-to-one 
correspondence with the ground states configuration 
allows us to specify the boundary conditions by the 
same index $b\in P^+_{r-n-1}$. 

In the presence of the boundary weight (\ref{eq:K-sol}), 
the $\sigma$-invariance (\ref{eq:s-inv}) is broken when 
we fix $a\in P^+_{r-n}$ at the right-most corner. 
In order to determine which 
$K(a+\bar{\varepsilon}_i , a, a|z)$ takes the largest 
among $K(a+\bar{\varepsilon}_\mu , a, a|z)$'s, let us 
consider the boundary Boltzmann weights 
in the `low temperature' limit $x\rightarrow 0$. 

Here we assume that the constant $\eta$ belongs to 
one of the following $n$ disjoint intavals 
$$
-\bar{a}_{n-1}<\eta <\frac{r-\bar{a}_0 -\bar{a}_{n-1}}{2}; 
~~~ 
-\frac{\bar{a}_{i-1}+\bar{a}_i}{2}<\eta 
<-\bar{a}_i ,\,\, (1\leqq i\leqq n-1). 
$$
Note that 
\begin{equation}
\bar{a}_{n-1} < \cdots <\bar{a}_1 <\bar{a}_0 
<r+\bar{a}_{n-1}, 
\label{eq:a-cond}
\end{equation}
for $a\in P^+_{r-n}$ \cite{JKMO}. 
We further restrict the spectral parameter $z$ to 
satisfy 
\begin{eqnarray}
-\dfrac{n}{2}<-\bar{a}_{n-1}-\eta <z<0, && 
\mbox{if \, $-\bar{a}_{n-1}<\eta <
\dfrac{r-\bar{a}_0 -\bar{a}_{n-1}}{2}$}, 
\label{eq:cond-A} \\
-\dfrac{n}{2}<\bar{a}_{i}+\eta <z<0, && 
\mbox{if \, $-\dfrac{\bar{a}_{i-1}+\bar{a}_i}{2}<\eta 
<-\bar{a}_i ,\,\, (1\leqq i\leqq n-1)$.}
\label{eq:cond-B}
\end{eqnarray}

When $-\bar{a}_{n-1}<\eta <
\frac{r-\bar{a}_0 -\bar{a}_{n-1}}{2}$, 
by using 
(\ref{eq:a-cond}) and (\ref{eq:cond-A}) we have 
$$
\begin{array}{ll}
\bar{a}_\mu +\eta -z>\bar{a}_\mu +\eta +z
\geqq \bar{a}_{n-1} +\eta +z>0, & 
(0\leqq \mu \leqq n-1), \\
r-\bar{a}_\mu -\eta -z>r-\bar{a}_\mu -\eta +z
\geqq r-\bar{a}_0 -\eta +z>0, & 
(0\leqq \mu \leqq n-1). 
\end{array}
$$
Thus the boundary Boltzmann weights behave like 
\begin{equation}
f_a (z)^{-1}K\left( \left. a+\bar{\varepsilon}_\mu 
\begin{array}{c} a \\ a' 
\end{array} \right| z \right)\sim 
\zeta^{\frac{2(\bar{a}_\mu +\eta )}{r}-1}, ~~ 
(x\rightarrow 0). 
\label{eq:A-behave}
\end{equation}
We therefore find from (\ref{eq:A-behave}) that 
$K(a+\bar{\varepsilon}_0 , a, a|z)$ takes the largest 
for $-\bar{a}_{n-1}<\eta <
\frac{r-\bar{a}_0 -\bar{a}_{n-1}}{2}$. 

When $-\frac{\bar{a}_{i-1}+\bar{a}_i}{2}<\eta 
<-\bar{a}_i$ $(1\leqq i\leqq n-1)$, by using 
(\ref{eq:a-cond}) and (\ref{eq:cond-B}) we have 
$$
\begin{array}{ll}
\bar{a}_\mu +\eta -z>\bar{a}_\mu +\eta +z
\geqq \bar{a}_{i-1} +\eta +z>0, & 
(0\leqq \mu \leqq i-1), \\
\bar{a}_\mu +\eta +z<\bar{a}_\mu +\eta -z
\leqq \bar{a}_{i} +\eta -z<0, & 
(i\leqq \mu \leqq n-1), \\
r-\bar{a}_\mu -\eta -z>r-\bar{a}_\mu -\eta +z
\geqq r-\bar{a}_0 -\eta +z>0, & 
(0\leqq \mu \leqq n-1). 
\end{array}
$$
Thus the boundary Boltzmann weights behave like 
\begin{equation}
f_a (z)^{-1}K\left( \left. a+\bar{\varepsilon}_\mu 
\begin{array}{c} a \\ a' 
\end{array} \right| z \right)\sim \left\{ 
\begin{array}{ll}
\zeta^{\frac{2(\bar{a}_\mu +\eta )}{r}-1}, & 
(0\leqq \mu \leqq i-1), \\
\zeta^{\frac{2(\bar{a}_\mu +\eta )}{r}+1}, & 
(i\leqq \mu \leqq n-1), \end{array} 
\right. ~~ (x\rightarrow 0). 
\label{eq:B-behave}
\end{equation}
We therefore find from (\ref{eq:B-behave}) that 
$K(a+\bar{\varepsilon}_i , a, a|z)$ takes the largest 
for $-\frac{\bar{a}_{i-1}+\bar{a}_i}{2}<\eta 
<-\bar{a}_i$ $(1\leqq i\leqq n-1)$. 

As is seen above, which 
$K(a+\bar{\varepsilon}_i , a, a|z)$ takes the largest 
depends on the value of $\eta$. 
Suppose that we fix $\eta$ such that 
$K(a+\bar{\varepsilon}_i , a, a|z)$ takes the largest 
value among 
$K(a+\bar{\varepsilon}_\mu , a, a|z)$'s. 
Then the boundary condition is be labeled by 
\begin{equation}
b=a-\omega_i \,\,\, (0\leqq i \leqq n-1). 
\label{eq:bc}
\end{equation}
For fixed $i$ in (\ref{eq:bc}) we should 
rather rewrite (\ref{eq:K-sol}) to the expression 
normalized by $K(a+\bar{\varepsilon}_i , a, a|z)$: 
\begin{equation}
K^{(i)}\left( \left. a+\bar{\varepsilon}_\mu 
\begin{array}{c} a \\ a' 
\end{array} \right| z \right)=f_a^{(i)}(z)
\dfrac{[\bar{a}_i +\eta -z]}{[\bar{a}_i +\eta +z]} 
\dfrac{[\bar{a}_\mu+\eta +z]}{[\bar{a}_\mu+\eta -z]} 
\delta_{aa'}. 
\label{eq:K-sol'}
\end{equation}

\section{Fusion of $K$-matrices} 

In order to determine the normalization factor $f_a^{(i)}(z)$ 
let us introduce the fusion of $K$-matrices. 
The fusion hierarchy of the boundary ABF model 
(boundary $A^{(1)}_1$-face model) was constructed 
in \cite{BPO}, and the fusion procedure of the boundary 
vertex models was considered in \cite{MN,HSY}. 

\subsection{Bulk and boundary face operators}

In this subsection we reformulate the bulk Boltamann 
weight $W$ and the boundary Boltzmann weight $K$ as 
elements of the bulk and boundary face operators. Let 
\begin{eqnarray*}
\Omega_z^{(b,a)}&=& \left\{ \begin{array}{ll}
\mathbb{C}v^{(b,a)} & 
\mbox{if $(a,b)\in P^2$ is admissible}, \\
0 & \mbox{otherwise}, 
\end{array} \right. \\
\Omega_z &=&\bigoplus_{a,b} \Omega_z^{(b,a)}. 
\end{eqnarray*}
Then the $W$-operator is defined as \cite{JKMO} 
\begin{equation}
W^{\Omega_{z_1}, \Omega_{z_{2}}}
(v^{(d,a)}\otimes v^{(c,d)})
=\sum_{b} v^{(c,b)}\otimes v^{(b,a)}
W\left( \left. \begin{array}{cc} 
c & d \\ b & a \end{array} \right| z_1 -z_2 \right). 
\end{equation}
Furthermore, if we introduce the $K$-operator by 
\begin{equation}
\begin{array}{rcl}
K(z): \Omega_z &\rightarrow& \Omega_{-z} \\
K(z)v^{(b,a)} &=& \displaystyle\sum_{a'} 
v^{(b,a')}K\left( \left. b 
\begin{array}{c} a \\ a' 
\end{array} \right| z \right), 
\end{array}
\end{equation}
the reflection equation (\ref{eq:faceRE}) can be regarded 
as the equality of linear operators: 
\begin{equation}
K_2 (z_2 ) W_{2\,1}^{\Omega_{z_2}, \Omega_{-z_{1}}}K_1 (z_1 )
W_{1\,2}^{\Omega_{z_1}, \Omega_{z_{2}}}=
W_{2\,1}^{\Omega_{-z_2}, \Omega_{-z_{1}}}K_1 (z_1 )
W_{1\,2}^{\Omega_{z_1}, \Omega_{-z_{2}}}K_2 (z_2 ), 
\label{eq:RE-int}
\end{equation}
where the subscripts of $W$'s and 
$K$'s denote the spaces on which they 
nontrivially act. The both sides of (\ref{eq:RE-int}) map 
$\Omega_{z_1}\otimes \Omega_{z_2}$ 
to $\Omega_{-z_1}\otimes \Omega_{-z_2}$. 

\subsection{Fusion procedure of $W$-operator}

For fixed $m$ ($2\leqq m\leqq n$) 
let $z_j =z+\frac{m+1}{2}-j$ ($1\leqq j\leqq m$), and 
for $0\leqq \mu_1 <\cdots <\mu_m \leqq n-1$ we denote 
$$
a_0 =a, ~~~~ 
a_j = a+\bar{\varepsilon}_{\mu_1}+ 
\cdots +\bar{\varepsilon}_{\mu_j}\,\,(1\leqq j\leqq m), 
$$
and 
$$
a_j^\sigma 
= a+\bar{\varepsilon}_{\mu_{\sigma(1)}}+ 
\cdots +\bar{\varepsilon}_{\mu_{\sigma(j)}}\,\,
(1\leqq j\leqq m), 
$$
for $\sigma\in\mathfrak{S}_m$. Note that 
$a_m^\sigma =a_m$. 
Let 
\begin{eqnarray*}
\wedge^m (\Omega_z^{(a_m , a_0)})
&=&\sum_{\sigma\in\mathfrak{S}_m} (\mbox{sgn}\,\sigma)\,
\Omega_{z_1}^{(a_1^\sigma , a_0)}\otimes 
\Omega_{z_2}^{(a_2^\sigma , a_1^\sigma )}\otimes \cdots 
\otimes \Omega_{z_m}^{(a_m , a_{m-1}^\sigma)}, \\
\wedge^m (\Omega_z )&=&\bigoplus_{a_0 , a_m} 
\wedge^m (\Omega_z^{(a_m , a_0)}), 
\end{eqnarray*} 
and let $v^{(a_m , a)}$ stand for the one dimensional 
basis of $\wedge^m (\Omega_z^{(a_m , a_0)})$. 
Then $W_{2\,1}^{\Omega_{z_2}, \Omega_{z_1}}$ is the 
fusion operator associated with 
$\wedge^2 (\Omega_z)$, because 
$\mbox{Im}\,(W_{2\,1}^{\Omega_{z_2}, \Omega_{z_1}}
|\Omega_{z_1}\otimes \Omega_{z_2})=\wedge^2 (\Omega_z)$. 
For general $m$, the fusion operator $\pi^{(m)}_\pm$ 
associated with $\wedge^m (\Omega_{\pm z})$'s are 
given as follows \cite{JKMO}: 
$$
\begin{array}{rcl}
\pi_+^{(m)}&=&W_{m\,m-1}^{\Omega_{z_m}, \Omega_{z_{m-1}}}\cdots 
W_{3\,2}^{\Omega_{z_3},\Omega_{z_2}}
W_{3\,1}^{\Omega_{z_3},\Omega_{z_1}}
W_{2\,1}^{\Omega_{z_2}, \Omega_{z_1}}, \\
\pi_-^{(m)}&=&W_{1\,2}^{\Omega_{-z_1}, \Omega_{-z_{2}}}
W_{1\,3}^{\Omega_{-z_1}, \Omega_{-z_{3}}}
W_{2\,3}^{\Omega_{-z_2}, \Omega_{-z_{3}}}\cdots 
W_{m-1\,m}^{\Omega_{-z_{m-1}}, \Omega_{-z_m}}. 
\end{array}
$$

The $m$-fold fuzed $W$-operator as an intertwiner 
on $\Omega_{z_1}\otimes \wedge^m 
(\Omega_{z_2})$ should be defined as 
\begin{equation}
W^{\Omega_{z_1}, \wedge^m (\Omega_{z_2})}
(v^{(d,a)}\otimes v^{(d_m , d)})=\sum_{a_m} 
v^{(d_m, a_m)}\otimes v^{(a_m , a)}
W^{(1,m)}\left( \left. 
\begin{array}{cc} d_m & d \\ a_m & a \end{array} \right| 
z_1 -z_2 \right), 
\end{equation}
where $W^{(1,m)}$'s are the horizontal $m$-fold fused 
Boltzmann weights, and $d_m -d=
\bar{\varepsilon}_{\lambda_1}+ \cdots 
+\bar{\varepsilon}_{\lambda_m}$ with $0\leqq \lambda_1 
< \cdots <\lambda_m \leqq n-1$. 
Another $m$-fold fuzed $W$-operator as an intertwiner 
on $\wedge^m (\Omega_{z_1})\otimes 
\Omega_{z_2}$ should be defined as 
\begin{equation}
W^{\wedge^m (\Omega_{z_1}),\Omega_{z_2}}
(v^{(a_m ,a)}\otimes v^{(b_m , a_m)})=\sum_{b} 
v^{(b_m, b)}\otimes v^{(b,a)}
W^{(m,1)}\left( \left. 
\begin{array}{cc} b_m & a_m \\ b & a \end{array} \right| 
z_1 -z_2 \right), 
\end{equation}
where $W^{(m,1)}$'s are the vertical $m$-fold fused 
Boltzmann weights, and $b_m -b=
\bar{\varepsilon}_{\lambda_1}+ \cdots 
+\bar{\varepsilon}_{\lambda_m}$ with $0\leqq \lambda_1 
< \cdots <\lambda_m \leqq n-1$. 
See Ref. \cite{AJMP} as for the definitions 
of horizontal and vertical fuzed Boltzmann weights. 
The results are also summarized in Appendix A of 
the present paper. 

Furthermore, we denote the dual space of $\Omega_z$ 
by $\Omega^*_z \cong \wedge^{n-1} (\Omega_z )$. Let 
\begin{eqnarray*}
\Omega_z^{*(b,a)}&=& \left\{ \begin{array}{ll}
\mathbb{C}v^{*(b,a)} 
& \mbox{if $(b,a)\in P^2$ is admissible}, \\
0 & \mbox{otherwise}, 
\end{array} \right. \\
\Omega^*_z &=&\bigoplus_{a,b} \Omega_z^{(b,a)}. 
\end{eqnarray*}
The dual $W$-operators are defined as 
\begin{equation}
\begin{array}{rcl}
W^{\Omega_{z_1}, \Omega^*_{z_{2}}}
(v^{(d,a)}\otimes v^{*(c,d)})
&=&\displaystyle\sum_{b} v^{(c,b)}\otimes v^{*(b,a)}
W^{\Omega_{z_1}, \Omega^*_{z_{2}}}
\left( \begin{array}{cc} 
c & d \\ b & a \end{array} \right). \\[6mm]
W^{\Omega^*_{z_1}, \Omega_{z_{2}}}
(v^{*(d,a)}\otimes v^{(c,d)})
&=&\displaystyle\sum_{b} v^{*(c,b)}\otimes v^{(b,a)}
W^{\Omega^*_{z_1}, \Omega_{z_{2}}}
\left( \begin{array}{cc} 
c & d \\ b & a \end{array} \right). \\[6mm]
W^{\Omega^*_{z_1}, \Omega^*_{z_{2}}}
(v^{*(d,a)}\otimes v^{*(c,d)})
&=&\displaystyle\sum_{b} v^{*(c,b)}\otimes v^{*(b,a)}
W^{\Omega^*_{z_1}, \Omega^*_{z_{2}}}
\left( \begin{array}{cc} 
c & d \\ b & a \end{array} \right). 
\end{array}
\end{equation}
The dual Boltzmann weights are graphically represented 
as follows: 

\unitlength 1mm
\begin{picture}(100,20)
\put(30,7){$W^{\Omega_{z_1}, \Omega^*_{z_{2}}}
\left( \begin{array}{cc} c & d 
\\ b & a \end{array} \right)=$} 
\put(73,3){\begin{picture}(101,0)
\multiput(10,0)(-1.0,0){10}{\circle*{0.5}}
\put(1,0){\vector(-1,0){1}}
\put(10,0){\vector(0,1){10}}
\put(0,0){\vector(0,1){10}}
\multiput(10,10)(-1.0,0){10}{\circle*{0.5}}
\put(1,10){\vector(-1,0){1}}
\multiput(-3,5)(2.2,0){7}{\line(1,0){1.2}}
\put(12.,5){\vector(1,0){1}}
\put(13.8,4.2){$z_1$}
\multiput(5,-3)(0,2.2){7}{\line(0,1){1.2}}
\put(5,12){\vector(0,1){1}}
\put(4,14){$z_2$}
\put(-2,10.5){$c$}
\put(10.5,10.1){$d$}
\put(10.5,-1.5){$a$}
\put(-2.3,-1.8){$b$}
\end{picture}
}
\end{picture}

\unitlength 1mm
\begin{picture}(100,20)
\put(30,7){$W^{\Omega^*_{z_1}, \Omega_{z_{2}}}
\left( \begin{array}{cc} c & d 
\\ b & a \end{array} \right)=$} 
\put(73,3){\begin{picture}(101,0)
\put(10,0){\vector(-1,0){10}}
\multiput(10,0)(0,1){10}{\circle*{0.5}}
\put(10,9){\vector(0,1){1}}
\multiput(0,0)(0,1){10}{\circle*{0.5}}
\put(0,9){\vector(0,1){1}}
\put(10,10){\vector(-1,0){10}}
\multiput(-3,5)(2.2,0){7}{\line(1,0){1.2}}
\put(12.,5){\vector(1,0){1}}
\put(13.8,4.2){$z_1$}
\multiput(5,-3)(0,2.2){7}{\line(0,1){1.2}}
\put(5,12){\vector(0,1){1}}
\put(4,14){$z_2$}
\put(-2,10.5){$c$}
\put(10.5,10.1){$d$}
\put(10.5,-1.5){$a$}
\put(-2.3,-1.8){$b$}
\end{picture}
}
\end{picture}

\unitlength 1mm
\begin{picture}(100,20)
\put(30,7){$W^{\Omega^*_{z_1}, \Omega^*_{z_{2}}}
\left( \begin{array}{cc} c & d 
\\ b & a \end{array} \right)=$} 
\put(73,3){\begin{picture}(101,0)
\multiput(10,0)(-1.0,0){10}{\circle*{0.5}}
\put(1,0){\vector(-1,0){1}}
\multiput(10,0)(0,1){10}{\circle*{0.5}}
\put(10,9){\vector(0,1){1}}
\multiput(0,0)(0,1){10}{\circle*{0.5}}
\put(0,9){\vector(0,1){1}}
\multiput(10,10)(-1.0,0){10}{\circle*{0.5}}
\put(1,10){\vector(-1,0){1}}
\multiput(-3,5)(2.2,0){7}{\line(1,0){1.2}}
\put(12.,5){\vector(1,0){1}}
\put(13.8,4.2){$z_1$}
\multiput(5,-3)(0,2.2){7}{\line(0,1){1.2}}
\put(5,12){\vector(0,1){1}}
\put(4,14){$z_2$}
\put(-2,10.5){$c$}
\put(10.5,10.1){$d$}
\put(10.5,-1.5){$a$}
\put(-2.3,-1.8){$b$}
\end{picture}
}
\end{picture}

\noindent Here, 
\unitlength 1mm
\begin{picture}(18,10)
\put(4,0.8){\begin{picture}(101,0)
\multiput(10,0)(-1.0,0){10}{\circle*{0.5}}
\put(1,0){\vector(-1,0){1}}
\put(11.2,-1){$a$}
\put(-2.5,-1){$b$}
\end{picture}
}
\end{picture}
implies that $(b,a)\in P^2$ is admissible. 

\subsection{Fusion procedure of $K$-operator}

Now we wish to construct the $m$-fold 
fusion of $K$-operator mapping $\wedge^m (\Omega_{z})$ 
to $\wedge^m (\Omega_{-z})$: 
\begin{equation}
K^{(m)}(z) v^{(a_m ,a)}=\sum_{a'} v^{(a_m ,a')}
K^{(m)}\left( \left. a_{m} 
\begin{array}{c} a \\ a' 
\end{array} \right| z \right). 
\label{eq:df-Km-op}
\end{equation}
Here we use the 
same notation $z_j$, $a_j$, $a_j^\sigma$ $(1\leqq j 
\leqq m)$ as in the previous subsection. 

Let us define the $m$-fold fusion of $K$ matrices 
in an inductive manner as follows. 
For $m=2$ let 
\begin{equation}
\begin{array}{cl}
&K_+^{(2)}\left( \left. a_2 
\begin{array}{c} a \\ a 
\end{array} \right| z \right) \\[5mm]
=&\displaystyle\sum_{
\sigma\in \mathfrak{S}_2} \mbox{sgn}\,\sigma\, 
K_1 \left( \left. a_1 \begin{array}{c} a \\ a 
\end{array} \right| z_1 \right)
W_{12}\left( \left. \begin{array}{cc} 
a_2 & a_1 \\ a_1^\sigma & a \end{array} 
\right| z_1 +z_2 \right)
K_2 \left( \left. a_1^\sigma \begin{array}{c} a \\ a 
\end{array} \right| z_2 \right) 
\end{array}
\label{eq:K2-fuse}
\end{equation}

\vspace{3mm} 

\unitlength 1mm
\begin{picture}(100,20)
\put(15.5,0){\begin{picture}(101,0)
\put(0,10){$=\,\,\,\,\displaystyle\sum_{
\sigma\in \mathfrak{S}_m} \mbox{sgn}\,\sigma $}
\end{picture}
}
\put(45,-3){\begin{picture}(101,0)
\put(0,0){\vector(0,1){10}}
\put(0,10){\vector(0,1){10}}
\put(20,20){\vector(-1,0){10}}
\put(10,20){\vector(-1,0){10}}
\put(10,10){\vector(0,1){10}}
\put(10,10){\vector(-1,0){10}}
\put(0,0){\line(1,1){20}}
\put(-1,-3){$a$}
\put(11,9){$a$}
\put(20.5,20){$a$}
\put(9,21.3){$a_1$}
\put(-4,9.5){$a_1^\sigma$}
\put(-2,21.3){$a_2$}
\multiput(-2,5)(2,0){3}{\line(1,0){1}}
\put(4,5){\vector(1,0){1}}
\put(-5,4.3){$z_2$}
\multiput(5,5)(0,2){8}{\line(0,1){1}}
\put(5,21){\vector(0,1){1}}
\put(2.5,22.8){\small$-z_2$}
\multiput(-2,15)(2,0){8}{\line(1,0){1}}
\put(14,15){\vector(1,0){1}}
\put(-5,14.3){$z_1$}
\multiput(15,15)(0,2){3}{\line(0,1){1}}
\put(15,21){\vector(0,1){1}}
\put(12.5,22.8){\small$-z_1$}
\end{picture}
}
\end{picture}

\vspace{8mm}

\noindent and for $m>2$ let 
\begin{equation}
\begin{array}{rcl}
K_+^{(m)}\left( \left. a_{m} 
\begin{array}{c} a \\ a 
\end{array} \right| z \right) 
&=&\displaystyle\sum_{
\sigma\in \mathfrak{S}_m} \mbox{sgn}\,\sigma\, 
K_+^{(m-1)}\left( \left. a_{m-1} \begin{array}{c} a \\ a 
\end{array} \right| z+\tfrac{1}{2} \right) \\[8mm]
&\times&
W^{(m-1,1)}\left( \left. \begin{array}{cc} 
a & a_{m-1} \\ a_1^\sigma & a \end{array} 
\right| z+\frac{1}{2}+z_m \right)
K \left( \left. a_1^\sigma \begin{array}{c} a \\ a 
\end{array} \right| z_m \right) \\[8mm]
&=&\displaystyle\sum_{
\sigma\in \mathfrak{S}_m} \mbox{sgn}\,\sigma\, 
K \left( \left. a_{1} \begin{array}{c} a \\ a 
\end{array} \right| z_1 \right) 
W^{(1,m-1)}\left( \left. \begin{array}{cc} 
a & a_{1} \\ a_{m-1}^\sigma & a \end{array} 
\right| z-\tfrac{1}{2}+z_1 \right) \\[8mm]
&\times&
K_+^{(m-1)}\left( \left. a_{m-1}^\sigma 
\begin{array}{c} a \\ a 
\end{array} \right| z-\frac{1}{2} \right)
\end{array}
\label{eq:K-rec}
\end{equation}

\vspace{8mm}

\unitlength 1mm
\begin{picture}(100,20)
\put(33,0){\begin{picture}(101,0)
\put(0,10){$=\,\,\,\,\displaystyle\sum_{
\sigma\in \mathfrak{S}_m} \mbox{sgn}\,\sigma $}
\end{picture}
}
\put(63,-3){\begin{picture}(101,0)
\put(0,0){\vector(0,1){10}}
\put(0,10){\vector(0,1){8}}
\put(0,18){\vector(0,1){10}}
\put(10,28){\vector(-1,0){10}}
\put(18,28){\vector(-1,0){8}}
\put(28,28){\vector(-1,0){10}}
\put(10,10){\vector(-1,0){10}}
\put(10,10){\vector(0,1){10}}
\put(18,18){\vector(-1,0){8}}
\put(18,18){\vector(0,1){10}}
\put(10,18){\vector(-1,0){10}}
\put(10,18){\vector(0,1){10}}
\put(0,0){\line(1,1){28}}
\put(-1,-3){$a$}
\put(11,9){$a$}
\put(18.8,17.3){$a$}
\put(28.5,28){$a$}
\put(17,29.3){$a_1$}
\put(7,29.3){$a_{m-1}$}
\put(-2,29.3){$a_{m}$}
\put(-4,9.5){$a_1^\sigma$}
\put(-8,17.5){$a_{m-1}^\sigma$}
\put(3,12.3){$\ddots$}
\put(12,21.2){$\ddots$}
\end{picture}
}
\end{picture}

\vspace{8mm}

\noindent Here $W^{(1,m-1)}$ and $W^{(m-1,1)}$ 
are horizontal and vertical fuzed Boltzmann weights, 
respectively. 

Successive application of the Yang-Baxter equation 
(\ref{eq:WYBE}) and the reflection equation 
(\ref{eq:faceRE}) indicate the reflection equations 
involving $K_+^{(m)}(z)$: 
\begin{equation}
\begin{array}{cl}
& K_2 (z_2 )W_{21}^{\Omega_{z_2},\wedge^m (\Omega_{-z_1})}
K^{(m)}_{+\,1}(z_1 )W_{12}^{\wedge^m (\Omega_{z_1}),\Omega_{z_2}} 
\\[3mm] =&
W_{21}^{\Omega_{-z_2},\wedge^m (\Omega_{-z_1})}
K^{(m)}_{+\,1} (z_1 )
W_{12}^{\wedge^m (\Omega_{z_1}),\Omega_{-z_2}}
K_2 (z_2 ), \\[3mm]
&K^{(m)}_{+\,2} (z_2 )
W_{21}^{\wedge^m (\Omega_{z_2}),\wedge^m (\Omega_{-z_1})}
K^{(m)}_{+\,1} (z_1 )
W_{12}^{\wedge^m (\Omega_{z_1}),\wedge^m (\Omega_{z_2})} 
\\[3mm] =&
W_{21}^{\wedge^m (\Omega_{-z_2}),\wedge^m (\Omega_{-z_1})}
K^{(m)}_{+\,1} (z_1 )
W_{12}^{\wedge^m (\Omega_{z_1}),\wedge^m (\Omega_{-z_2})}
K^{(m)}_{+\,2} (z_2 ). 
\end{array}
\label{eq:fuseRE}
\end{equation}
The both sides of the first one of eqs. (\ref{eq:fuseRE}) 
map $\wedge^m (\Omega_{z_1})\otimes \Omega_{z_2}$ to 
$\wedge^m (\Omega_{-z_1})\otimes \Omega_{-z_2}$, while 
those of the second one map 
$\wedge^m (\Omega_{z_1})\otimes \wedge^m (\Omega_{z_2})$ 
to 
$\wedge^m (\Omega_{-z_1})\otimes \wedge^m (\Omega_{-z_2})$. 

Another fused $K$-matrix is defined in an inductive manner 
as follows. For $m=2$ let 
\begin{equation}
\begin{array}{cl}
&K_-^{(2)}\left( \left. a_2 
\begin{array}{c} a \\ a 
\end{array} \right| z \right) \\[5mm]
=&\displaystyle\sum_{
\sigma\in \mathfrak{S}_2} \mbox{sgn}\,\sigma\, 
K_2 \left( \left. a_1 \begin{array}{c} a \\ a 
\end{array} \right| z_2 \right)
W_{21}\left( \left. \begin{array}{cc} 
a_2 & a_1 \\ a_1^\sigma & a \end{array} 
\right| z_1 +z_2 \right)
K_1 \left( \left. a_1^\sigma \begin{array}{c} a \\ a 
\end{array} \right| z_1 \right), 
\end{array}
\label{eq:K2'-fuse}
\end{equation}
and for $m>2$ let 
\begin{equation}
\begin{array}{rcl}
K_-^{(m)}\left( \left. a_{m} 
\begin{array}{c} a \\ a 
\end{array} \right| z \right) 
&=&\displaystyle\sum_{
\sigma\in \mathfrak{S}_m} \mbox{sgn}\,\sigma\, 
K_+^{(m-1)}\left( \left. a_{m-1} \begin{array}{c} a \\ a 
\end{array} \right| z-\tfrac{1}{2} \right) \\[8mm]
&\times&
W^{(m-1,1)}\left( \left. \begin{array}{cc} 
a & a_{m-1} \\ a_1^\sigma & a \end{array} 
\right| z-\tfrac{1}{2}+z_1 \right)
K \left( \left. a_1^\sigma \begin{array}{c} a \\ a 
\end{array} \right| z_1 \right) \\[8mm]
&=&\displaystyle\sum_{
\sigma\in \mathfrak{S}_m} \mbox{sgn}\,\sigma\, 
K \left( \left. a_{1} \begin{array}{c} a \\ a 
\end{array} \right| z_m \right) 
W^{(1,m-1)}\left( \left. \begin{array}{cc} 
a & a_{1} \\ a_{m-1}^\sigma & a \end{array} 
\right| z+\tfrac{1}{2}+z_m \right) \\[8mm]
&\times&
K_+^{(m-1)}\left( \left. a_{m-1}^\sigma 
\begin{array}{c} a \\ a 
\end{array} \right| z+\tfrac{1}{2} \right). 
\end{array}
\label{eq:K'-rec}
\end{equation}

Two kinds of $m$-fold fused $K$-matrices are related 
as follows: 
\begin{equation}
K_+^{(m)}(z)\pi_+^{(m)}=\pi_-^{(m)}K_-^{(m)}(z), 
\label{eq:Kpi}
\end{equation}
Note that (\ref{eq:Kpi}) with $m=2$ directly 
follows from the reflection equation (\ref{eq:faceRE}), 
and also that (\ref{eq:Kpi}) with $m>2$ follows from 
the Yang-Baxter equation (\ref{eq:WYBE}) and the 
reflection equation (\ref{eq:faceRE}). 
The relation (\ref{eq:Kpi}) implies that 
$\mbox{Im}\,(K_+^{(m)}(z)|\wedge^m (\Omega_{z}))
=\wedge^m (\Omega_{-z})$, and hence $K_+^{(m)}(z)$ 
can be regarded as a bona-fide $K$-matrix mapping 
$\wedge^m (\Omega_{z})$ to $\wedge^m (\Omega_{-z})$ 
satisfying (\ref{eq:fuseRE}). 

\subsection{Boundary crossing symmetry}

By taking account of the recursion relation 
(\ref{eq:K-rec}), the explicit expression of 
$K_+^{(m)}(z)$ can be obtained as follows: 
\begin{equation}
\begin{array}{cl}
&K_+^{(m)}\left( \left. a+\displaystyle\sum_{j=1}^m 
\bar{\varepsilon}_{\mu_j} 
\begin{array}{c} a \\ a 
\end{array} \right| z \right) \\[9mm] =&(-1)^{{}_mC_2}
\displaystyle\prod_{j<k} r_1 (z_j +z_k ) 
\prod_{j=1}^{m-1} \dfrac{[2z-j]}{[2z+j]} 
\prod_{j=1}^{m} f_a^{(i)} (z_j ) 
\dfrac{[\bar{a}_{i}+\eta -z_j ]}
{[\bar{a}_{i}+\eta +z_j ]}
\dfrac{[\bar{a}_{\mu_j}+\eta +z_1 ]}
{[\bar{a}_{\mu_j}+\eta -z_n ]}. 
\end{array}
\label{eq:K_+^m}
\end{equation}
Note that the sign factor $(-1)^{{}_mC_2}$ results from 
the permutation $(1, \cdots , m)\mapsto (m, \cdots , 1)$. 
See the figure below eq. (\ref{eq:K-rec}). 

Since $\wedge^n (\Omega_z^{(a,a)})\cong \mathbb{C}$, 
the $n$-fold fused $K$-matrix should be a scalar. 
By putting the scalar to be equal to unity 
we obtain the normalization factor $f_a^{(i)} (z)$ 
of $K$-matrix in (\ref{eq:K-sol'}) as follows: 
\begin{equation}
f_a^{(i)}(z) =\zeta^{\frac{n-1}{n}\frac{r-1}{r}-
\frac{2\bar{a}_i}{r}}
\dfrac{g(\zeta )p_a^{(i)}(\zeta )p_a^{(i)}(x^2 \zeta^{-1})}
{g(\zeta^{-1})p_a^{(i)}(\zeta^{-1})p_a^{(i)}(x^2 \zeta )}, 
\label{eq:f_a^i}
\end{equation}
where $\zeta =x^{2z}$, and 
$$
\begin{array}{c}
g(\zeta )=\dfrac{(x^{2n+2}\zeta^2 ; x^{4n}, x^{2r})_\infty 
(x^{2(r+n-1)}\zeta^2 ; x^{4n}, x^{2r})_\infty }
{(x^{2r}\zeta^2 ; x^{4n}, x^{2r})_\infty 
(x^{4n}\zeta^2 ; x^{4n}, x^{2r})_\infty }, \\[6mm]
p_a^{(i)}(\zeta )=\displaystyle\prod_{j=0}^{n-1} 
\dfrac{(x^{2(\bar{a}_i +\eta +j)}\zeta 
; x^{2n}, x^{2r})_\infty }
{(x^{2(r+n-j-1-\bar{a}_i -\eta )}\zeta ; x^{2n}, x^{2r})_\infty }
\dfrac{(x^{2(r-\bar{a}_j -\eta )}\zeta 
; x^{2n}, x^{2r})_\infty }
{(x^{2(\bar{a}_j +\eta +n-1)}\zeta ; x^{2n}, x^{2r})_\infty }. 
\end{array}
$$

Let us define the dual $K$-operator and 
its matrix element as follows: 
\begin{equation}
K^* (z) v^{*(b,a)} =\sum_{a'} v^{*(b,a')}
K^* \left( \left. b 
\begin{array}{c} a \\ a' 
\end{array} \right| z \right). 
\end{equation}
When $m=n-1$ we identify the dual $K$-matrix with 
the fused $K$-matrix as follows:  
\begin{equation}
K^* \left( \left. a-\bar{\varepsilon}_{\mu} 
\begin{array}{c} a \\ a' 
\end{array} \right| z \right) =
K_+^{(n-1)}\left( \left. a+\displaystyle\sum_{\nu =1 
\atop \nu\neq \mu}^n 
\bar{\varepsilon}_{\nu} 
\begin{array}{c} a \\ a' 
\end{array} \right| z \right). 
\label{eq:K-dual}
\end{equation}
For lator convenience we further introduce the 
$\hat{K}$-operator 
\begin{equation}
\hat{K}(z) v^{(b,a)} =\sum_{b'} v^{(b',a)} 
K^* \left( \left. a 
\begin{array}{c} b \\ b' 
\end{array} \right| -z-\frac{n}{2} \right). 
\end{equation}

The boundary crossing symmetry can be obtained by 
substituting (\ref{eq:K-dual}) into 
(\ref{eq:K-rec}) with $m=n$
\begin{equation}
K^* \left( \left. b 
\begin{array}{c} a \\ a' 
\end{array} \right| z \right) =\delta_{aa'} 
\sum_{d} \dfrac{G_d}{G_a} W\left( \left. \begin{array}{cc} 
d & a \\ a & b \end{array} \right| 2z \right) 
K\left( \left. d 
\begin{array}{c} a \\ a 
\end{array} \right| -\dfrac{n}{2}-z \right). 
\label{eq:b-cross}
\end{equation}
The boundary crossing relations were found in 
\cite{Skl,HSY} for vertex-type models, and 
in \cite{BPO,BFKZ} for face-type models. As for 
$A^{(1)}_1$, $B^{(1)}_n$, $C^{(1)}_n$, $D^{(1)}_n$ 
and $A^{(2)}_n$-face model cases, see \cite{BFKZ}. 
This is the first to derive the boundary crossing 
relation for the boundary $A^{(1)}_{n-1}$-face model 
with $n>2$. The vertex-type version of (\ref{eq:b-cross}) 
is given by (4.12) in \cite{bBela}. 

\section{Correlation functions and difference equations}

\subsection{Vertex operators and commutation relations} 

For $b=a-\omega_i$ ($1\leqq i\leqq n$), let 
${\cal H}_{l,k}$ be the space of admissible paths 
$(\cdots , a_2 , a_1 , a_0)$ such that 
\begin{equation}
a_0 =a, ~~~ a_j -a_{j-1}\in \left\{ 
\bar{\varepsilon}_0 , \bar{\varepsilon}_1 , 
\cdots , \bar{\varepsilon}_{n-1} 
\right\}, \mbox{ for $j=1, 2, 3, \cdots$, }~~~~ 
a_j=b+\omega_{j+i} \mbox{ for 
$j\gg 1$}, 
\end{equation}
where $k=a+\rho , l=b+\rho$. 

Following \cite{FJMMN,AJMP} we can identify 
the type I vertex operators with the half 
transfer matrices. Here we need four types of 
vertex operators: 

\unitlength 1mm
\begin{picture}(140,100)(0,30)
\put(-5,60){\begin{picture}(101,0)
\put(30,55){\vector(0,1){10}}
\put(30,55){\vector(-1,0){10}}
\put(30,65){\vector(-1,0){10}}
\put(40,55){\vector(0,1){10}}
\put(40,55){\vector(-1,0){10}}
\put(40,65){\vector(-1,0){10}}
\put(50,55){\vector(0,1){10}}
\put(50,55){\vector(-1,0){10}}
\put(50,65){\vector(-1,0){10}}
\put(60,55){\vector(0,1){10}}
\put(60,55){\vector(-1,0){10}}
\put(60,65){\vector(-1,0){10}}
\put(59,53){$a$}
\put(58,66){$a\!+\!\bar{\varepsilon}_\mu$}
\multiput(16,60)(2,0){23}{\line(1,0){1}}
\put(62,60){\vector(1,0){1.5}}
\multiput(35,53)(0,2){7}{\line(0,1){1}}
\multiput(45,53)(0,2){7}{\line(0,1){1}}
\multiput(55,53)(0,2){7}{\line(0,1){1}}
\put(35,67){\vector(0,1){1.5}}
\put(45,67){\vector(0,1){1.5}}
\put(55,67){\vector(0,1){1.5}}
\put(34,51){$z_2$}
\put(44,51){$z_2$}
\put(54,51){$z_2$}
\put(64,59.5){$z_1$}
\put(70,59){$=\phi_b^{(a+\bar{\varepsilon}_\mu ,a)}
(z_1 -z_2 ): 
{\cal H}_{l,k}\longrightarrow 
{\cal H}_{l,k+\bar{\varepsilon}_\mu},$}
\end{picture}
}
\put(-5,35){\begin{picture}(101,0)
\put(30,65){\vector(0,-1){10}}
\put(30,55){\vector(-1,0){10}}
\put(30,65){\vector(-1,0){10}}
\put(40,65){\vector(0,-1){10}}
\put(40,55){\vector(-1,0){10}}
\put(40,65){\vector(-1,0){10}}
\put(50,65){\vector(0,-1){10}}
\put(50,55){\vector(-1,0){10}}
\put(50,65){\vector(-1,0){10}}
\put(60,65){\vector(0,-1){10}}
\put(60,55){\vector(-1,0){10}}
\put(60,65){\vector(-1,0){10}}
\put(59,53){$a$}
\put(58,66){$a\!-\!\bar{\varepsilon}_\mu$}
\multiput(18,60)(2,0){23}{\line(1,0){1}}
\put(18,60){\vector(-1,0){1.5}}
\multiput(35,53)(0,2){7}{\line(0,1){1}}
\multiput(45,53)(0,2){7}{\line(0,1){1}}
\multiput(55,53)(0,2){7}{\line(0,1){1}}
\put(35,67){\vector(0,1){1.5}}
\put(45,67){\vector(0,1){1.5}}
\put(55,67){\vector(0,1){1.5}}
\put(34,51){$z_2$}
\put(44,51){$z_2$}
\put(54,51){$z_2$}
\put(64,59.5){$z_1$}
\put(70,59){$=\phi^{b}_{(a-\bar{\varepsilon}_\mu ,a)}
(z_2 -z_1 ): 
{\cal H}_{l,k}\longrightarrow 
{\cal H}_{l,k-\bar{\varepsilon}_\mu}$}
\end{picture}
}
\put(-5,10){\begin{picture}(101,0)
\multiput(30,55)(0,1){10}{\circle*{0.5}}
\put(30,64){\vector(0,1){1}}
\put(30,55){\vector(-1,0){10}}
\put(30,65){\vector(-1,0){10}}
\multiput(40,55)(0,1){10}{\circle*{0.5}}
\put(40,64){\vector(0,1){1}}
\put(40,55){\vector(-1,0){10}}
\put(40,65){\vector(-1,0){10}}
\multiput(50,55)(0,1){10}{\circle*{0.5}}
\put(50,64){\vector(0,1){1}}
\put(50,55){\vector(-1,0){10}}
\put(50,65){\vector(-1,0){10}}
\multiput(60,55)(0,1){10}{\circle*{0.5}}
\put(60,64){\vector(0,1){1}}
\put(60,55){\vector(-1,0){10}}
\put(60,65){\vector(-1,0){10}}
\put(59,53){$a$}
\put(58,66){$a\!-\!\bar{\varepsilon}_\mu$}
\multiput(16,60)(2,0){23}{\line(1,0){1}}
\put(62,60){\vector(1,0){1.5}}
\multiput(35,53)(0,2){7}{\line(0,1){1}}
\multiput(45,53)(0,2){7}{\line(0,1){1}}
\multiput(55,53)(0,2){7}{\line(0,1){1}}
\put(35,67){\vector(0,1){1.5}}
\put(45,67){\vector(0,1){1.5}}
\put(55,67){\vector(0,1){1.5}}
\put(34,51){$z_2$}
\put(44,51){$z_2$}
\put(54,51){$z_2$}
\put(64,59.5){$z_1$}
\put(70,59){$=\phi_b^{*(a-\bar{\varepsilon}_\mu ,a)}
(z_1 -z_2 ): 
{\cal H}_{l,k}\longrightarrow 
{\cal H}_{l,k-\bar{\varepsilon}_\mu},$}
\end{picture}
}
\put(-5,-15){\begin{picture}(101,0)
\multiput(30,65)(0,-1){10}{\circle*{0.5}}
\put(30,56){\vector(0,-1){1}}
\put(30,55){\vector(-1,0){10}}
\put(30,65){\vector(-1,0){10}}
\multiput(40,65)(0,-1){10}{\circle*{0.5}}
\put(40,56){\vector(0,-1){1}}
\put(40,55){\vector(-1,0){10}}
\put(40,65){\vector(-1,0){10}}
\multiput(50,65)(0,-1){10}{\circle*{0.5}}
\put(50,56){\vector(0,-1){1}}
\put(50,55){\vector(-1,0){10}}
\put(50,65){\vector(-1,0){10}}
\multiput(60,65)(0,-1){10}{\circle*{0.5}}
\put(60,56){\vector(0,-1){1}}
\put(60,55){\vector(-1,0){10}}
\put(60,65){\vector(-1,0){10}}
\put(59,53){$a$}
\put(58,66){$a\!+\!\bar{\varepsilon}_\mu$}
\multiput(18,60)(2,0){23}{\line(1,0){1}}
\put(18,60){\vector(-1,0){2}}
\multiput(35,53)(0,2){7}{\line(0,1){1}}
\multiput(45,53)(0,2){7}{\line(0,1){1}}
\multiput(55,53)(0,2){7}{\line(0,1){1}}
\put(35,67){\vector(0,1){1.5}}
\put(45,67){\vector(0,1){1.5}}
\put(55,67){\vector(0,1){1.5}}
\put(34,51){$z_2$}
\put(44,51){$z_2$}
\put(54,51){$z_2$}
\put(64,59.5){$z_1$}
\put(70,59){$=\phi^{*b}_{(a+\bar{\varepsilon}_\mu ,a)}
(z_2 -z_1 ): 
{\cal H}_{l,k}\longrightarrow 
{\cal H}_{l,k+\bar{\varepsilon}_\mu}$. }
\end{picture}
}
\end{picture}

\noindent In what follows 
we often suppress the letter $b$ specifying 
the boundary condition. 

It follows from the Yang-Baxter equation (\ref{eq:WYBE}) 
and the boundary condition that these vertex operators 
satisfy the following commutation relations: 
\begin{equation}
\begin{array}{rcl}
{\phi}^{(c,d)}(z_2)\phi^{(d,a)}(z_1)
&=&\displaystyle\sum_{d}W\left(\left.
\begin{array}{cc}
c & d \\
b & a \end{array}\right|z_1-z_2 \right)
{\phi}^{(c,d)}(z_1)
{\phi}^{(d,a)}(z_2), \\[7mm]
{\phi}^{*(c,b)}(z_2)\phi^{(b,a)}(z_1)
&=&\displaystyle\sum_{d}
W^{\Omega_{z_1},\Omega^*_{z_2}}\left( 
\begin{array}{cc} c & d \\ 
b & a \end{array} \right)
{\phi}^{(c,d)}(z_1)
{\phi}^{*(d,a)}(z_2), \\[7mm]
{\phi}^{(c,b)}(z_2)\phi^{*(b,a)}(z_1)
&=&\displaystyle\sum_{d}
W^{\Omega^*_{z_1},\Omega_{z_2}}\left( 
\begin{array}{cc} c & d \\ 
b & a \end{array} \right)
{\phi}^{*(c,d)}(z_1)
{\phi}^{(d,a)}(z_2), \\[7mm]
{\phi}^{*(c,b)}(z_2)\phi^{*(b,a)}(z_1)
&=&\displaystyle\sum_{d}
W^{\Omega^*_{z_1},\Omega^*_{z_2}}\left( 
\begin{array}{cc} c & d \\ 
b & a \end{array} \right)
{\phi}^{*(c,d)}(z_1)
{\phi}^{*(d,a)}(z_2). 
\end{array}
\label{eq:CR-VO}
\end{equation}
Here $W^{\Omega_{z_1},\Omega^*_{z_2}}$, 
$W^{\Omega^*_{z_1},\Omega_{z_2}}$ and 
$W^{\Omega^*_{z_1},\Omega^*_{z_2}}$ denote 
the dual $W$-operators. See Appendix A 
concerning their matrix elements. 

Furtermore, the unitarity relations with respect to 
the $W$-operators imply the inversion relation of 
the vertex operators: 
\begin{equation}
\sum_{\mu =0}^{n-1} {\phi}_{(a,a+\bar{\varepsilon}_\mu )}(-z)
{\phi}^{(a+\bar{\varepsilon}_\mu ,a)}(z)=1. 
\label{eq:V-uni}
\end{equation}
Thanks to the crossing symmetry with respect to 
$W$-operators (see Appendix A) 
we also have the duality identities: 
\begin{equation}
{\phi}^{*(a-\bar{\varepsilon}_\mu ,a)}(z)
=G_a^{-1} 
{\phi}_{(a-\bar{\varepsilon}_\mu ,a)}(-z-\tfrac{n}{2}), 
~~~~
{\phi}^*_{(a+\bar{\varepsilon}_\mu ,a)}(z)
=G_a 
{\phi}^{(a+\bar{\varepsilon}_\mu ,a)}(-z-\tfrac{n}{2}). 
\label{eq:V-dual}
\end{equation}

Using the vertex operators introduced in the previous 
section, the transfer matrix for the semi-infinite 
lattice is expressed as follows: 
\begin{equation}
T^{(i)}_B (z) 
=\displaystyle\sum_{\mu =0}^{n-1} 
\phi_{(a, a+\bar{\varepsilon}_\mu )}(z)
K^{(i)}\left( \left. a+\bar{\varepsilon}_\mu 
\begin{array}{c} a \\ a 
\end{array} \right| z \right) 
\phi^{(a+\bar{\varepsilon}_\mu ,a)}(z) 
\label{eq:b-trans}
\end{equation}

\unitlength 1mm
\begin{picture}(140,35)(0,0)
\put(38,5){\begin{picture}(101,0)
\put(3,14){$=\displaystyle\sum_{b}$}
\put(30,5){\vector(0,1){10}}
\put(30,25){\vector(0,-1){10}}
\put(30,5){\vector(-1,0){10}}
\put(30,15){\vector(-1,0){10}}
\put(30,25){\vector(-1,0){10}}
\put(40,5){\vector(0,1){10}}
\put(40,25){\vector(0,-1){10}}
\put(40,5){\vector(-1,0){10}}
\put(40,15){\vector(-1,0){10}}
\put(40,25){\vector(-1,0){10}}
\put(50,5){\vector(0,1){10}}
\put(50,25){\vector(0,-1){10}}
\put(50,5){\vector(-1,0){10}}
\put(50,15){\vector(-1,0){10}}
\put(50,25){\vector(-1,0){10}}
\put(60,5){\vector(0,1){10}}
\put(60,25){\vector(0,-1){10}}
\put(60,5){\vector(-1,0){10}}
\put(60,15){\vector(-1,0){10}}
\put(60,25){\vector(-1,0){10}}
\put(59,3){$a$}
\put(62,14){$b$}
\put(59,26){$a$}
\put(70,5){\vector(-1,1){10}}
\put(70,25){\vector(-1,-1){10}}
\put(70,5){\line(0,1){20}}
\multiput(60,5)(2.2,0){5}{\line(1,0){1.2}}
\multiput(60,25)(2.2,0){5}{\line(1,0){1.2}}
\put(69,3){$a$}
\put(69,26){$a$}
\multiput(18,10)(2,0){24}{\line(1,0){1}}
\put(15,9.5){$z_1$}
\multiput(65,10)(1.0,1.0){4}{\circle*{0.5}}
\put(69,14){\vector(1,1){1}}
\multiput(20,20)(2,0){23}{\line(1,0){1}}
\put(14,19.5){$-z_1$}
\multiput(70,15)(-1.0,1.0){4}{\circle*{0.5}}
\put(66,19){\vector(-1,1){1}}
\end{picture}
}
\end{picture}

\noindent From (\ref{eq:b-trans}) and (\ref{eq:V-dual}) 
we also have another expression
\begin{equation}
T^{(i)}_B (z) 
=\displaystyle\sum_{\mu= 0}^{n-1} G_{a+
\bar{\varepsilon}_\mu}
\phi^{*(a,a+\bar{\varepsilon}_\mu )}(z)
K^{(i)}\left( \left. a+\bar{\varepsilon}_\mu 
\begin{array}{c} a \\ a 
\end{array} \right| z \right) 
\phi^{(a+\bar{\varepsilon}_\mu ,a)}(z). 
\end{equation}

\subsection{Derivation of difference equations} 

In sections 2 and 3 we fix the normalization of 
$W$ and $K$ such that the maximal eigenvalues of 
the boundary transfer matrix $T^{(i)}_B (z)$ is 
equal to unity in the thermodynamic limit. Thus 
the boundary vacuum state $|k-\omega_i , k\rangle _B$ 
in ${\cal H}_{k-\omega_i , k}$ and its dual 
${}_B\langle k-\omega_i , k|$ in 
${\cal H}^*_{k-\omega_i , k}$ should satisfy 
\begin{equation}
\begin{array}{rcl}
K^{(i)}\left( \left. a+\bar{\varepsilon}_\mu 
\begin{array}{c} a \\ a 
\end{array} \right| z \right) 
\phi^{(a+\bar{\varepsilon}_\mu ,a)}(z) 
|k-\omega_i , k\rangle _B 
&=&\phi^{(a+\bar{\varepsilon}_\mu ,a)}(-z) 
|k-\omega_i , k\rangle _B , \\ 
{}_B\langle k-\omega_i , k| 
\phi^*_{(a+\bar{\varepsilon}_\mu ,a)}(z)
K^{(i)}\left( \left. a+\bar{\varepsilon}_\mu 
\begin{array}{c} a \\ a 
\end{array} \right| z \right) 
&=&{}_B\langle k-\omega_i , k| 
\phi^*_{(a+\bar{\varepsilon}_\mu ,a)}(-z). 
\label{eq:Ref-Pro}
\end{array}
\end{equation}

Define the $(N+1)$-point correlation function as 
\begin{equation}
\begin{array}{cl}
&F^{(i)}(z_1 , z_2 , \cdots , z_N )^{a_0 , a_{1}, 
a_2 , \cdots , a_{N-1} , a_N} \\
=& {}_B\langle k-\omega_i , k| 
\phi^{(a_0 ,a_{1})}(z_1 )\phi^{(a_1 ,a_{2})}(z_2 )
\cdots \phi^{(a_{N-1},a_{N})}(z_N )
|k-\omega_i , k\rangle _B , 
\label{eq:N-pt}
\end{array}
\end{equation}
where we assume that $N\equiv 0$ mod $n$ for simplicity. 
It follows from (\ref{eq:CR-VO}), (\ref{eq:Ref-Pro}) 
and (\ref{eq:V-dual}) 
that correlation functions should satisfy 

\noindent 1. $R$-matrix symmery: 
\begin{equation}
\begin{array}{cl}
&F^{(i)}(\cdots , z_{j+1} , z_{j}, \cdots )^{\cdots , 
a_{j-1},a_j ,a_{j+1} \cdots}\\
=&\displaystyle\sum_{a'_j} W\left(\left.
\begin{array}{cc}
a_{j+1} & a_j \\
a'_j & a_{j-1} \end{array}\right|z_1 -z_2 \right)
F^{(i)}(\cdots , z_{j} , z_{j+1}, \cdots )^{\cdots , 
a_{j-1},a'_j ,a_{j+1} \cdots}, 
\end{array}
\label{eq:R-symm}
\end{equation}
2. Reflection properties: 
\begin{eqnarray}
F^{(i)}(\cdots , -z_{N})^{\cdots , 
a_{N-1}, a_N}
&=&K^{(i)}\left( \left. a_{N-1}
\begin{array}{c} a_N \\ a_N 
\end{array} \right| z \right) 
F^{(i)}(\cdots , z_{N})^{\cdots , 
a_{N-1}, a_N}, 
\\
F^{(i)}(-z_{1} -\frac{n}{2}, \cdots , )^{
a_{0}, a_1, \cdots}
&=&\hat{K}^{(i)}\left( \left. a_{1}
\begin{array}{c} a_0 \\ a_0 
\end{array} \right| z \right) 
F^{(i)}(z_{1}, \cdots )^{a_{0}, a_1, \cdots}. 
\label{eq:Ref-symm}
\end{eqnarray}
These relations can be shown by 
the same discussion as in \cite{JKKMW,bBela}. 

Let 
$$
\begin{array}{cl}
&F^{(i)}(z_1 , z_2 , \cdots , z_N ) \\
=&\displaystyle\sum_{a_0 , a_{1}, \cdots , a_N} 
v^{(a_0, a_1)}\otimes v^{(a_1, a_2)}\otimes 
\cdots \otimes v^{(a_{N-1}, a_N)}
F^{(i)}(z_1 , z_2 , \cdots , z_N )^{a_0 , a_{1}, a_2 , 
\cdots , a_{N-1} , a_N}. 
\end{array}
$$
Then we conclude from (\ref{eq:R-symm}--\ref{eq:Ref-symm}) 
that that the $\Omega_{z_1} \otimes \cdots \otimes 
\Omega_{z_N}$-valued correlation function 
$F^{(i)}(z_1 , \cdots , z_N )$ should satisfy 
the following difference equations 
\begin{equation}
\begin{array}{rcl}
T_j F^{(i)}_N (z, z'| z_1, \cdots , z_N ) 
&=& W_{j j-1}^{\Omega_{z_j -n},\Omega_{z_{j-1}}}
\cdots W_{j 1}^{\Omega_{z_j -n},\Omega_{z_{1}}}
\hat{K}_j(-z_j ) \\[2mm]
&\times&
W_{1 j}^{\Omega_{z_{1},\Omega_{-z_j}}}\cdots
W_{j-1 j}^{\Omega_{z_{j-1},\Omega_{-z_j}}}
W_{j+1 j}^{\Omega_{z_{j+1},\Omega_{-z_j}}} \cdots
W_{N j}^{\Omega_{z_{N},\Omega_{-z_j}}} \\[2mm]
&\times& K_j (z_j ) 
W_{j N}^{\Omega_{z_j},\Omega_{z_{N}}} \cdots
W_{j j+1}^{\Omega_{z_j},\Omega_{z_{j+1}}}
F^{(i)}_N (z, z'| z_1, \cdots , z_N ), 
\label{eq:N-diff}
\end{array}
\end{equation}
where 
$$
T_j f (z, z'| z_1, \cdots , z_j , 
\cdots , z_N ) =f(z, z'| z_1, \cdots , 
z_j -n, \cdots , z_N ). 
$$

\subsection{Simple difference equations}

In this subsection we consider the correlation functions 
of the form 
\begin{equation}
\begin{array}{rcl}
P_i^{(a+\bar{\varepsilon}_\mu , a)} (z_1 , z_2 )
&=&{}_B\langle k-\omega_i , k| 
\phi_{(a,a+\bar{\varepsilon}_\mu )}(-z_1 ) 
\phi^{(a+\bar{\varepsilon}_\mu ,a)}(z_2 )
|k-\omega_i , k\rangle _B \\[2mm] 
&=&G_{a+\bar{\varepsilon}_\mu} \times 
{}_B\langle k-\omega_i , k| 
\phi^{*(a,a+\bar{\varepsilon}_\mu )}(-z_1 ) 
\phi^{(a+\bar{\varepsilon}_\mu ,a)}(z_2 )
|k-\omega_i , k\rangle _B . 
\end{array}
\label{eq:1-pt}
\end{equation}
You can show by the similar way as in (\ref{eq:N-diff}) 
that this correlation function satisfy the following 
difference equations: 
\begin{equation}
\begin{array}{rcl}
T_1 P_i^{(a+\bar{\varepsilon}_\lambda , a)} (z_1 , z_2 )
&=& \displaystyle\sum_{\mu \nu} 
\dfrac{G_{a+\bar{\varepsilon}_\lambda} 
G_{a-\bar{\varepsilon}_\mu}}{G_a^2} 
K^{(i)}\left( \left. a+\bar{\varepsilon}_\lambda 
\begin{array}{c} a \\ a 
\end{array} \right| z_1-n \right) \\[6mm]
&\times&W\left( \left. \begin{array}{cc} 
a+\bar{\varepsilon}_\lambda & a \\
a & a- \bar{\varepsilon}_\mu 
\end{array} \right| -z_1 -z_2 \right)
K^{*(i)}\left( \left. a-\bar{\varepsilon}_\mu 
\begin{array}{c} a \\ a 
\end{array} \right| z_1-\tfrac{n}{2} \right) \\[6mm]
&\times&W\left( \left. \begin{array}{cc} 
a+\bar{\varepsilon}_\nu & a \\
a & a- \bar{\varepsilon}_\mu 
\end{array} \right| z_2 -z_1 \right)
P_i^{(a+\bar{\varepsilon}_\nu , a)} (z_1 , z_2 ), 
\end{array}
\label{eq:T_1}
\end{equation}
\begin{equation}
\begin{array}{rcl}
T_2 P_i^{(a+\bar{\varepsilon}_\lambda , a)} (z_1 , z_2 )
&=& \displaystyle\sum_{\mu \nu} 
\dfrac{G_{a+\bar{\varepsilon}_\lambda} 
G_{a-\bar{\varepsilon}_\mu}}{G_a^2} 
W\left( \left. \begin{array}{cc} 
a+\bar{\varepsilon}_\lambda & a \\
a & a- \bar{\varepsilon}_\mu 
\end{array} \right| z_1 -z_2 \right)
\\[6mm]
&\times&K^{*(i)}\left( \left. a-\bar{\varepsilon}_\mu 
\begin{array}{c} a \\ a 
\end{array} \right| z_2-\tfrac{n}{2} \right)
W\left( \left. \begin{array}{cc} 
a+\bar{\varepsilon}_\nu & a \\
a & a- \bar{\varepsilon}_\mu 
\end{array} \right| -z_1 -z_2 \right) \\[6mm]
&\times&K^{*(i)}\left( \left. a+\bar{\varepsilon}_\nu 
\begin{array}{c} a \\ a 
\end{array} \right| z_1-n \right) 
P_i^{(a+\bar{\varepsilon}_\nu , a)} (z_1 , z_2 ). 
\end{array}
\label{eq:T_2}
\end{equation}
Set 
$$
P_i^{(a)} (z_1 , z_2 ) =
\sum_{\lambda =0}^{n-1} 
P_i^{(a+\bar{\varepsilon}_\lambda , a)} (z_1 , z_2 ), 
$$
and we restrict ourselves to the limiting case 
that $K$-matrix is a certain scalar as done in 
\cite{JKKMW,bBela}. 
Then from (\ref{eq:T_1},\ref{eq:T_2}) the difference 
equations for $P_i^{(a)} (z_1 , z_2 )$ are given as 
\begin{equation}
\begin{array}{rcl}
\dfrac{T_1 P_i^{(a)} (z_1 , z_2 )}
{P_i^{(a)} (z_1 , z_2 )}&=&(x^{-2n}\zeta_1^2 )^{
\frac{n-1}{r}}
r_1 (2z_1 -n) r_1 (-z_+ ) \dfrac{[-z_+ +n]}{[-z_+ +1]} 
r_1 (-z_- )\dfrac{[-z_- +n]}{[-z_- +1]}, \\[5mm]
\dfrac{T_2 P_i^{(a)} (z_1 , z_2 )}
{P_i^{(a)} (z_1 , z_2 )}&=&(x^{-2n}\zeta_2^2 )^{
\frac{n-1}{r}}
r_1 (2z_2 -n) r_1 (-z_+ ) \dfrac{[-z_+ +n]}{[-z_+ +1]} 
r_1 (z_- )\dfrac{[z_- +n]}{[z_- +1]}, 
\end{array}
\label{eq:LSP}
\end{equation}
where we use the notation $z_\pm =z_1 \pm z_2$, and 
also use the following sum formulae: 
\begin{equation}
\begin{array}{rcl}
\displaystyle\sum_{\lambda =0}^{n-1} 
\dfrac{G_{a+\bar{\varepsilon}_\lambda}}{G_a}
W\left( \left. \begin{array}{cc} 
a+\bar{\varepsilon}_\lambda & a \\
a & a- \bar{\varepsilon}_\mu 
\end{array} \right| z \right) &=& 
r_1 (z) \dfrac{[z+n]}{[z+1]}, \\[6mm]
\displaystyle\sum_{\mu =0}^{n-1} 
\dfrac{G_{a-\bar{\varepsilon}_\mu}}{G_a}
W\left( \left. \begin{array}{cc} 
a+\bar{\varepsilon}_\nu & a \\
a & a- \bar{\varepsilon}_\mu 
\end{array} \right| z \right) &
=&r_1 (z) \dfrac{[z+n]}{[z+1]}. 
\end{array}
\label{eq:sum-formula}
\end{equation}
The both formulae (\ref{eq:sum-formula}) are 
equivalent to 
\begin{equation}
\prod_{\nu =0\atop \nu\neq \mu}^{n-1} 
\frac{[a_{\mu\nu}+1]}{[a_{\mu\nu}]}+
\sum_{\lambda =0\atop \lambda\neq \mu}^{n-1}
\frac{[z+a_{\lambda\mu}+1][1]}
{[z+1][a_{\lambda\mu}]}
\prod_{\nu =0\atop \nu\neq \lambda, \mu}^{n-1} 
\frac{[a_{\lambda\nu}+1]}{[a_{\lambda\nu}]}
=\frac{[z+n]}{[z+1]}. 
\label{eq:proof-sum}
\end{equation}
Note that the LHS of (\ref{eq:proof-sum}) is not 
singular at $\bar{a}_i =\bar{a}_j$ ($0\leqq i<j 
\leqq n-1$), and also that the LHS vanishes at 
$z=-n$. Hence the LHS is equal to the RHS up to 
a constant $C$. In order to show that $C=1$, it is 
sufficient to set $z=0$ and 
$$
a_{\mu\nu} =\left\{ \begin{array}{ll} 
n+\nu -\mu & (\nu =0, 1, \cdots , \mu -1 ), \\
\nu -\mu & (\nu =\mu +1, \cdots , n-1). 
\end{array} \right. 
$$
Thus we find from (\ref{eq:LSP}) and 
(\ref{eq:sum-formula}) that the correlation function 
has the form
\begin{equation}
P_i^{(a)} (z_1 , z_2 ) =C_i^{(a)} 
A(z_1 )A(z_2 )B(z_+ )B(z_-), 
\end{equation}
where $C_i^{(a)}$ is a constant, and 
$$
A(z)=a(\zeta^2 )a(\zeta^{-2}), ~~~~
a(\zeta )=
\frac{(x^{4n+2r-2}\zeta ; x^{2n}, x^{4n}, x^{2r})_\infty 
(x^{2n+2}\zeta ; x^{2n}, x^{4n}, x^{2r})_\infty }
{(x^{2n+2r}\zeta ; x^{2n}, x^{4n}, x^{2r})_\infty 
(x^{4n}\zeta ; x^{2n}, x^{4n}, x^{2r})_\infty }, 
$$
$$
B(z)=b(\zeta )b(\zeta^{-1}), ~~~~
b(\zeta )=
\frac{(x^{2r}\zeta ; x^{2n}, x^{2n}, x^{2r})_\infty 
(x^{4n}\zeta ; x^{2n}, x^{2n}, x^{2r})_\infty }
{(x^{2n+2r-2}\zeta ; x^{2n}, x^{2n}, x^{2r})_\infty 
(x^{2n+2}\zeta ; x^{2n}, x^{2n}, x^{2r})_\infty }. 
$$

\section{Concluding remarks} 

There are two ways in order to proceed further. 
The one is solving the difference equations 
(\ref{eq:N-diff}) to obtain the corresponding 
local state probabilities, 
while the other is constructing the boundary 
vacuum states in terms of bosonized vertex operators 
to do the same thing. 
Let us discuss the latter way here. 

Consider the bosons
$B_m^j\,(1\leqq j \leqq n-1, m \in \mathbb{Z}
\backslash \{0\})$
with the commutation relations \cite{FF,AKOS}
$$
[B_m^j,B_{m'}^k]
=\left\{ \begin{array}{l} 
m\dfrac{[(n-1)m]_x}{[nm]_x}
\dfrac{[(r-1)m]_x}{[rm]_x}\delta_{m+m',0},~(j=k)\\[6mm]
-mx^{{\rm sgn}(j-k)nm}\dfrac{[m]_x}{[nm]_x}
\dfrac{[(r-1)m]_x}{[rm]_x}\delta_{m+m',0},~(j\neq k), 
\end{array} \right. 
$$
where the symbol $[a]_x$ stands for
$(x^a-x^{-a})/(x-x^{-1})$.
Define $B_m^n$ by
\begin{eqnarray*}
\sum_{j=1}^n x^{-2jm}B_m^j=0.
\end{eqnarray*}
Using these oscillators $B_m^j$'s the bosonization of 
the type I vertex operators 
for the $A^{(1)}_{n-1}$-face model was given in 
\cite{AJMP} on the Fock space ${\cal F}_{l,k}$. 
Furthermore, we make the Ansatz \cite{JKKKM,MW} 
such that the boundary vacuum states and their dual 
have the form 
$$
|k-\omega_i , k\rangle _B =\exp 
(F_a^{(i)})|k-\omega_i , k\rangle , ~~~~
{}_B \langle k-\omega_i , k|=
{}_B \langle k-\omega_i , k|\exp (G_a^{(i)}), 
$$
where $|l , k\rangle $ is the highest 
weight of ${\cal F}_{l,k}$, and 
$$
\begin{array}{rcl}
F_a^{(i)}&=&-\displaystyle\frac{1}{2}\sum_{m>0} 
\sum_{s,t=1}^{n-1} 
\alpha_m^{st} B_{-m}^s B_{-m}^t +
\sum_{m>0} \sum_{t=1}^{n-1} \beta^{t,(i)}_{m,a} B_{-m}^t , \\
G_a^{(i)}&=&-\displaystyle\frac{1}{2}\sum_{m>0} 
\sum_{s,t=1}^{n-1} 
\gamma_m^{st} B_{-m}^s B_{-m}^t +
\sum_{m>0} \sum_{t=1}^{n-1} \delta^{t,(i)}_{m,a} B_{-m}^t . 
\end{array}
$$
Since ${\cal H}_{l,k}$ and ${\cal F}_{l,k}$ have 
different characters, the bosonized expressions of 
correlation functions can not be identified with the 
one defined in (\ref{eq:N-pt}). In order to obtain 
the correct bosonized formulae for correlation 
functions, we have to construct the BRST cohomology 
of appropriate complex which realizes the space of 
physical states ${\cal H}_{l,k}$ as subquotients of 
${\cal F}_{l,k}$'s. We wish to address this problem 
in a separate paper. 

\section*{Acknowledgement}

The author would like to thank K. Hasegawa and 
M. Okado for useful discussion. 

\appendix

\section{Appendix A. Fused $A^{(1)}_{n-1}$ Boltzmann weight}

Let us introduce the fusion of Boltzmann weights $W$ \cite{AJMP}. 

Let $\Lambda =\{ \lambda_1 , \cdots , \lambda_m \}$ 
be a subset of $N=\{ 0, 1, \cdots , n\}$ 
such that $\lambda_1 <\cdots <\lambda_m$. 
For $\kappa , \mu \in N$ 
set $\mu =\kappa$ if $\kappa \in \Lambda$, 
otherwise set $\mu \in \Lambda\cup \{\kappa \}$. 
For given $\kappa , \mu, \Lambda$ 
let $0\leqq \nu_1 <\cdots < \nu_m \leqq n-1$ 
be such that $\bar{\varepsilon}_\mu 
+\bar{\varepsilon}_{\nu_1}+ 
\cdots +\bar{\varepsilon}_{\nu_m}=
\bar{\varepsilon}_\kappa +
\bar{\varepsilon}_{\lambda_1}+ 
\cdots +\bar{\varepsilon}_{\lambda_m}$. 

The fusion of $W$ in the horizontal 
direction is constructed as follows. 
Let $a, b, d=d_0 , d_1 , 
\cdots d_{m-1}, d_{m}=c \in P$ satisfy 
$$
c=b+\bar{\varepsilon}_\mu , \,\,
d_{j}-d_{j-1}=\bar{\varepsilon}_{\lambda_j}\,
(1\leqq j\leqq m), \,\,
d=a+\bar{\varepsilon}_\kappa . 
$$
Note that $b=a+\bar{\varepsilon}_{\nu_1}+ 
\cdots +\bar{\varepsilon}_{\nu_m}$ from 
the definition of $\nu_j$'s. Let $\sigma 
\in \mathfrak{S}_m$ be a permutation of 
$(1, \cdots , m)$, and set 
$$
a_0^\sigma =a,\,\, a^{\sigma}_j =b^{\sigma}_{j-1}+
\bar{\varepsilon}_{\nu_{\sigma (j)}}\,
(1\leqq j\leqq m),\,\, a_m^\sigma =b. 
$$
Then $m$-fold anti-symmetric fusion of $W$ 
in the horizontal direction is given as 
\begin{equation}
W^{(1,m)}\left( \left. 
\begin{array}{cc} c & d \\ b & a \end{array} \right| 
z \right)=\sum_{\sigma \in \mathfrak{S}_m} 
\mbox{sgn}\,\sigma \prod_{j=1}^m 
W\left( \left. 
\begin{array}{cc} d_{j} & d_{j-1} \\ 
a_{j}^\sigma & a_{j-1}^\sigma \end{array} \right| 
z+\tfrac{m+1}{2}-j \right). 
\end{equation}
Note that $W^{(1,m)}$ is anti-symmetric 
with respect to $(\lambda_1 , \cdots , \lambda_m )$. 

Next consider the fusion in the vertical direction. 
We use the same $\kappa, \mu$, $\lambda_j$'s and 
$\nu_j$'s as before. Now we set 
$$
b=a+\bar{\varepsilon}_\mu , \,\,
a_{j}-a_{j-1}=\bar{\varepsilon}_{\lambda_j}\,
(1\leqq j\leqq m), \,\,
c=d+\bar{\varepsilon}_\kappa , 
$$
where $a_0 =a, a_m =d$. We have 
$c=b+\bar{\varepsilon}_{\nu_1}+ 
\cdots +\bar{\varepsilon}_{\nu_m}$. 
For $\sigma\in\mathfrak{S}_m$ set
$$
b_0^\sigma =b,\,\, b^{\sigma}_j =
b^{\sigma}_{j-1}+
\bar{\varepsilon}_{\nu_{\sigma (j)}}\,
(1\leqq j\leqq m),\,\, b_m^\sigma =c. 
$$
Then $m$-fold anti-symmetric fusion of $W$ 
in the vertical direction is given as 
\begin{equation}
W^{(m,1)}\left( \left. 
\begin{array}{cc} c & d \\ b & a \end{array} \right| 
v \right)=\sum_{\sigma \in \mathfrak{S}_m} 
\mbox{sgn}\,\sigma \prod_{j=1}^m 
W\left( \left. 
\begin{array}{cc} b_{j}^\sigma & a_{j} \\ 
b_{j-1}^\sigma & a_{j-1} \end{array} \right| 
z-\tfrac{m+1}{2}+j \right). 
\end{equation}
Note that $W^{(m,1)}$ is anti-symmetric 
with respect to $(\lambda_1 , \cdots , \lambda_m )$. 

We further introduce the fusion of $W$ 
in both horizontal and vertical directions. 
Let $0\leqq \kappa_1 < \cdots < \kappa_m \leqq n-1$, 
$0\leqq \mu_1 <\cdots <\mu_m \leqq n-1$, 
$0\leqq \lambda_1 < \cdots < \lambda_{m'}\leqq n-1$ and 
$0\leqq \nu_1 <\cdots <\nu_{m'}\leqq n-1$ satisfy 
$$
\sum_{j=1}^m \bar{\varepsilon}_{\kappa_j}+
\sum_{j=1}^{m'} \bar{\varepsilon}_{\lambda_j}=
\sum_{j=1}^m \bar{\varepsilon}_{\mu_j}+
\sum_{j=1}^{m'} \bar{\varepsilon}_{\nu_j}. 
$$
Let $a, b, c, d\in P$ satisfy 
$$
d=a+\sum_{j=1}^m \bar{\varepsilon}_{\kappa_j}, \, 
c=d+\sum_{j=1}^{m'} \bar{\varepsilon}_{\lambda_j}, \, 
b=a+\sum_{j=1}^{m'} \bar{\varepsilon}_{\nu_j}, \, 
c=b+\sum_{j=1}^m \bar{\varepsilon}_{\mu_j}. 
$$
Then the $m\times m'$--fold fusion of $W$ 
is defiend as the antisymmetrized product of 
the $m'$-fold fusion of $W$ in the 
horizontal direction: 
\begin{equation}
W^{(m,m')}\left( \left. 
\begin{array}{cc} c & d \\ b & a \end{array} \right| 
v \right)=\sum_{\sigma \in \mathfrak{S}_m} 
\mbox{sgn}\,\sigma \prod_{j=1}^m 
W_{II}^{(1,m')}\left( \left. 
\begin{array}{cc} b_{j}^\sigma & a_{j} \\ 
b_{j-1}^\sigma & a_{j-1} \end{array} \right| 
z-\tfrac{m+1}{2}+j \right), 
\end{equation}
where 
$$
b_0^\sigma =b,\,\, b^{\sigma}_j =
b^{\sigma}_{j-1}+
\bar{\varepsilon}_{\mu_{\sigma (j)}}\,
(1\leqq j\leqq m),\,\, b_m^\sigma =c. 
$$
The $W^{(m,m')}$ can be also 
defiend as the antisymmetrized product of 
the $m$-fold fusion of $W$ in the 
vertical direction: 
\begin{equation}
W^{(m,m')}\left( \left. 
\begin{array}{cc} c & d \\ b & a \end{array} \right| 
z \right)=\sum_{\sigma \in \mathfrak{S}_{m'}} 
\mbox{sgn}\,\sigma \prod_{j=1}^{m'} 
W^{(m,1)}\left( \left. 
\begin{array}{cc} d_{j} & d_{j-1} \\ 
a_{j}^\sigma & a_{j-1}^\sigma \end{array} \right| 
z+\tfrac{m'+1}{2}-j \right), 
\end{equation}
where 
$$
a_0^\sigma =a,\,\, a^{\sigma}_j =a^{\sigma}_{j-1}+
\bar{\varepsilon}_{\nu_{\sigma (j)}}\,
(1\leqq j\leqq m'),\,\, a_{m'}^\sigma =b. 
$$

The explicit expression of fused Boltzmann weights 
in both the horizontal and vertical directions are 
given as follows \cite{AJMP}: 

\begin{equation}
\begin{array}{rcl}
W^{(1,m)}\left( \left. 
\begin{array}{cc} a+\bar{\varepsilon}_\lambda 
+\bar{\varepsilon}_{\Lambda} 
& a+\bar{\varepsilon}_\lambda \\ 
a+\bar{\varepsilon}_{\Lambda} 
& a \end{array} \right| z \right) &=& 
(-1)^{m-1} r_m (z) 
\dfrac{\mbox{[}z+\frac{m-1}{2}\mbox{]}}
{\mbox{[}z+\frac{m+1}{2}\mbox{]}}
\displaystyle\prod_{j=1}^m 
\dfrac{\mbox{[}a_{\lambda\lambda_j}-1\mbox{]}}
{\mbox{[}a_{\lambda\lambda_j}\mbox{]}}, \\[8mm]
W^{(1,m)}\left( \left. 
\begin{array}{cc} a+\bar{\varepsilon}_\lambda 
+\bar{\varepsilon}_{\Lambda} 
& a+\bar{\varepsilon}_{\lambda_1} \\ 
a+\bar{\varepsilon}_{\Lambda} 
& a \end{array} \right| z \right) &=& 
(-1)^{m-1} r_m (z) 
\dfrac{\mbox{[}z+\frac{m-1}{2}+a_{\lambda\lambda_1}\mbox{]}
[1]}
{\mbox{[}z+\frac{m+1}{2}\mbox{]}[a_{\lambda\lambda_1}]}
\displaystyle\prod_{j=2}^m 
\dfrac{\mbox{[}a_{\lambda\lambda_j}-1\mbox{]}}
{\mbox{[}a_{\lambda\lambda_j}\mbox{]}}, \\[8mm]
W^{(1,m)}\left( \left. 
\begin{array}{cc} a+\bar{\varepsilon}_{\lambda_1} 
+\bar{\varepsilon}_{\Lambda} 
& a+\bar{\varepsilon}_{\lambda_1} \\ 
a+\bar{\varepsilon}_{\Lambda} 
& a \end{array} \right| z \right) &=& 
(-1)^{m-1} r_m (z) 
\displaystyle\prod_{j=2}^m 
\dfrac{\mbox{[}a_{\lambda_1\lambda_j}\mbox{]}}
{\mbox{[}a_{\lambda_1\lambda_j}+1\mbox{]}}, \\[8mm]
W^{(m,1)}\left( \left. 
\begin{array}{cc} a+\bar{\varepsilon}_\lambda 
+\bar{\varepsilon}_{\Lambda} 
& a+\bar{\varepsilon}_\Lambda \\ 
a+\bar{\varepsilon}_{\lambda} 
& a \end{array} \right| z \right) &=& 
(-1)^{m-1} r_m (z) 
\dfrac{\mbox{[}z+\frac{m-1}{2}\mbox{]}}
{\mbox{[}z+\frac{m+1}{2}\mbox{]}}
\displaystyle\prod_{j=1}^m 
\dfrac{\mbox{[}a_{\lambda_j\lambda}-1\mbox{]}}
{\mbox{[}a_{\lambda_j\lambda}\mbox{]}}, \\[8mm]
W^{(m,1)}\left( \left. 
\begin{array}{cc} a+\bar{\varepsilon}_\lambda 
+\bar{\varepsilon}_{\Lambda} 
& a+\bar{\varepsilon}_{\Lambda} \\ 
a+\bar{\varepsilon}_{\lambda_m} 
& a \end{array} \right| z \right) &=& 
(-1)^{m-1} r_m (z) 
\dfrac{\mbox{[}z+\frac{m-1}{2}+a_{\lambda\lambda_m}\mbox{]}
[1]}
{\mbox{[}z+\frac{m+1}{2}\mbox{]}[a_{\lambda\lambda_m}]}
\displaystyle\prod_{j=1}^{m-1} 
\dfrac{\mbox{[}a_{\lambda_j\lambda_m}-1\mbox{]}}
{\mbox{[}a_{\lambda_j\lambda_m}\mbox{]}}, \\[8mm]
W^{(m,1)}\left( \left. 
\begin{array}{cc} a+\bar{\varepsilon}_{\lambda_m} 
+\bar{\varepsilon}_{\Lambda} 
& a+\bar{\varepsilon}_{\Lambda} \\ 
a+\bar{\varepsilon}_{\lambda_m} 
& a \end{array} \right| z \right) &=& 
(-1)^{m-1} r_m (z) 
\displaystyle\prod_{j=1}^{m-1} 
\dfrac{\mbox{[}a_{\lambda_j\lambda_m}-1\mbox{]}}
{\mbox{[}a_{\lambda_j\lambda_m}\mbox{]}}. 
\end{array}
\end{equation}
Here we denote $\bar{\varepsilon}_{\lambda_1}+\cdots 
+\bar{\varepsilon}_{\lambda_m}$ by 
$\bar{\varepsilon}_{\Lambda}$ for simplicity. 

When $m=n-1$ we identify $W^{\Omega,\Omega^*}$ 
(resp. $W^{\Omega^*,\Omega}$) 
with $W^{(1,n-1)}$ (resp. $W^{(n-1,1)}$) as follows: 
\begin{equation}
W^{\Omega_{z_1},\Omega^*_{z_2}}\left( 
\begin{array}{cc} c & d \\ 
b & a \end{array} \right)=
(-1)^{n-1+\lambda +\nu}W^{(1,n-1)}\left( \left. 
\begin{array}{cc} c & d \\ b & a \end{array} \right| 
z_1 -z_2 \right), 
\end{equation}
where $b-a=-\bar{\varepsilon}_\nu$, 
$c-d=-\bar{\varepsilon}_\lambda$, 
and $(b,c)$, $(a,d)$ are admissible; and 
\begin{equation}
W^{\Omega^*_{z_1},\Omega_{z_2}}\left( 
\begin{array}{cc} c & d \\ 
b & a \end{array} \right)=(-1)^{n-1+\mu +\kappa} 
W^{(n-1,1)}\left( \left. 
\begin{array}{cc} c & d \\ b & a \end{array} \right| 
z_1 -z_2 \right), 
\end{equation}
where $d-a=-\bar{\varepsilon}_\kappa$, 
$c-b=-\bar{\varepsilon}_\mu$, and $(a,b)$, $(d,c)$ 
are admissible. 

The crossing symmetries are as follows: 
\begin{equation}
W^{\Omega_{z_1},\Omega^*_{z_2}}\left( 
\begin{array}{cc} c & d \\ 
b & a \end{array} \right)=
\dfrac{G_b}{G_c} W\left( \left. 
\begin{array}{cc} d & a \\ c & b \end{array} \right| 
z_2 -z_1 -\frac{n}{2} \right), 
\end{equation}
and 
\begin{equation}
W^{\Omega^*_{z_1},\Omega_{z_2}}\left( 
\begin{array}{cc} c & d \\ 
b & a \end{array} \right)=
\dfrac{G_b}{G_a} W\left( \left. 
\begin{array}{cc} b & c \\ a & d \end{array} \right| 
z_2 -z_1 -\frac{n}{2} \right). 
\end{equation}

\section{Appendix B. Proof of the reflection equation}

The aim of this Appendix is to give a simple sketch of 
the proof of the claim that (\ref{eq:K-sol}) solves the 
reflection equation (\ref{eq:faceRE}). Since 
$K$-matrix is diagonal in the sence that $K(a,b,c|z)
=0$ unless $b=c$, it is sufficient 
to consider the case $a=e=g$ in (\ref{eq:faceRE}). 
In this case the reflection equation reduces to 
\begin{equation}
\begin{array}{cl}
&\displaystyle\sum_{d} 
K\left( \left. d \begin{array}{c} a \\ a 
\end{array} \right| z_1 \right) 
K\left( \left. f \begin{array}{c} a \\ a 
\end{array} \right| z_2 \right)
W\left( \left. \! \begin{array}{cc} 
c & f \\ d & a \end{array} \right| z_1 +z_2 \right)
W\left( \left. \! \begin{array}{cc} 
c & d \\ b & a \end{array} \right| z_1 -z_2 \right) 
\\[7mm]
=& \displaystyle\sum_{d} 
K\left( \left. d \begin{array}{c} a \\ a 
\end{array} \right| z_1 \right)
K\left( \left. b \begin{array}{c} a \\ a 
\end{array} \right| z_2 \right)
W\left( \left. \! \begin{array}{cc} 
c & f \\ d & a \end{array} \right| z_1 -z_2 \right)
W\left( \left. \! \begin{array}{cc} 
c & d \\ b & a \end{array} \right| z_1 +z_2 \right). 
\end{array}
\label{eq:faceRE'}
\end{equation}
Note that (\ref{eq:faceRE'}) holds as $0=0$ 
unless the quartet $(a,b,c,f)$ is admissible. 
Assume that $(a,b,c,f)$ is admissible. Then 
there are the following three cases: 

\vspace{3mm}

\begin{tabular}{cl}
(i) & $b=f=a+\bar{\varepsilon}_\mu$, 
$c=a+2\bar{\varepsilon}_\mu$; \\
(ii) & $b=f=a+\bar{\varepsilon}_\mu$, 
$c=a+\bar{\varepsilon}_\mu +\bar{\varepsilon}_\nu$ 
($\mu \neq \nu$); \\
(iii) & $b=a+\bar{\varepsilon}_\mu$, 
$f=a+\bar{\varepsilon}_\nu$, 
$c=a+\bar{\varepsilon}_\mu +\bar{\varepsilon}_\nu$ 
($\mu \neq \nu$). 
\end{tabular}

\vspace{3mm}

\noindent For the case (i) the equation 
(\ref{eq:faceRE'}) is trivial because the only 
non-zero terms of both sides result from 
$d=a+\bar{\varepsilon}_\mu$. It is also easy to 
prove the case (ii). Up to now we did 
not use the explicit form of $K$-matrix (\ref{eq:K-sol}) 
except for its diagonal property. 

Let us prove the case (iii). By substituting 
(\ref{eq:BW}) and (\ref{eq:K-sol}) into 
(\ref{eq:faceRE'}), the reflection equation 
(\ref{eq:faceRE}) is equivalent to 
$$
\begin{array}{cl}
&\dfrac{[\bar{a}_\nu+\eta +z_2]}{[\bar{a}_\nu+\eta -z_2]} 
\left( 
\dfrac{[\bar{a}_\mu+\eta +z_1]}{[\bar{a}_\mu+\eta -z_1]} 
[z_1 +z_2 ][a_{\mu\nu}-z_1 +z_2]
+\dfrac{[\bar{a}_\nu+\eta +z_1]}{[\bar{a}_\nu+\eta -z_1]} 
[z_1 -z_2 ][a_{\mu\nu}+z_1 +z_2] \right)\\[5mm]
=&\dfrac{[\bar{a}_\mu+\eta +z_2]}{[\bar{a}_\mu+\eta -z_2]} 
\left( 
\dfrac{[\bar{a}_\mu+\eta +z_1]}{[\bar{a}_\mu+\eta -z_1]} 
[z_1 -z_2 ][a_{\mu\nu}-z_1 -z_2]
+\dfrac{[\bar{a}_\nu+\eta +z_1]}{[\bar{a}_\nu+\eta -z_1]} 
[z_1 +z_2 ][a_{\mu\nu}+z_1 -z_2] \right). 
\end{array}
$$
Let $f$ stand for the difference of both sides as the 
function of $z_1$. It is easy to show that the poles at 
$z_1 =\bar{a}_\mu+\eta$ and $z_2 =
\bar{a}_\nu+\eta $ are spurious. Thus the function $f$ 
is an entire function of $z_1$ with the quasi double 
periodicities. Suppose that $f$ is not identically zero. 
Then the transformation properties of $f$ contradict 
the positions of zeros at $z_1 =\pm z_2$. We therefore 
obtain $f=0$ and conclude that (\ref{eq:K-sol}) solves 
the reflection equation (\ref{eq:faceRE}).


\begin{thebibliography}{99}
\bibitem{JM} Jimbo M and Miwa T: 
{\it Algebraic analysis of solvable lattice models, 
CBMS Regional Conferences Series in Mathematics}, 1994. 
\bibitem{FJMMN}Foda O, Jimbo M and Miwa T, Miki K 
and Nakayashiki A: Vertex operators in solvable 
lattice models, {\it J. Math. Phys.} {\bf 35} (1994) 
13--46. 
\bibitem{ZZ}Zamolodchikov A B and Zamolodchikov Al B: 
Factorized $S$-matrices in two dimensions at the 
exact solutions of certain relativistic quantum 
field theory models, 
{\it Ann. Phys.} {\bf 120} (1979) 253--291. 
\bibitem{ESM}Baxter R J: {\it Exactly Solved Models 
in Statistical Mechanics, 
Academic Press, London}, 1982.
\bibitem{RE}Cherednik I V: Factorizing particles 
on a half-line and root systems,
{\it Theor. Math. Phys.} {\bf 61} (1984) 977--983.
\bibitem{bKZ}Cherednik I V: Degenerate affine Hecke 
algebras and two-dimensional particles, 
Proceedings of the RIMS Research Project 1991, 
``Infinite Analysis'' 
{\it Int. J. Mod. Phys.} {\bf A7} 
Suppl. {\bf 1A} (1992) 109--140. 
\bibitem{Skl}Sklyanin E K:
Boundary conditions for integrable quantum systems,
{\it J. Phys. A: Math. Gen.} {\bf 21} (1988) 2375--2389.
\bibitem{GZ}Ghoshal S and Zamolodchikov A B:
Boundary S-matrix and boundary state in two
dimensional integrable quantum field theory,
{\it Int. J. Mod. Phys.} {\bf A9} (1994) 3841--3886;
Erratum ibid. 4353.
\bibitem{Kul}Kulish P P: Yang--Baxter equation and 
reflection equations in integrable models, in {\it 
Low-dimensional models in statistial physics and 
quantum field theory}, Grosse H and Pittner L Eds., 
(Springer--Verlag, Berlin, 1996) 125--144. 
\bibitem{BPO}Behrend R E, Pearce P A and O'Brien D L: 
Interaction-Round-a-Face models with fixed boundary 
conditions: The ABF fusion hierarchy, 
{\it J. Stat. Phys.} {\bf 84} (1996) 1-48. 
\bibitem{bBela}Quano Y.-H: Difference equations for 
correlation functions of Belavin's 
$\mathbb{Z}_n$-symmetric model with boundary reflection, 
{\it J. Phys. A: Math Gen.} {\bf 33} (2000) 8275--8303. 
\bibitem{JKKMW}Jimbo M, Kedem R, Konno K, Miwa T and
Weston R: Difference equations in spin chains with a boundary,
{\it Nucl. Phys.} {\bf B448 [FS]} (1995) 429--456.
\bibitem{SPn}Quano Y.-H: Spontaneous polarization of the 
$\mathbb{Z}_n$-Baxter model, {\it Mod. Phys. Lett.} 
{\bf A8} (1993) 3363--3375. 
\bibitem{JMO}Jimbo M, Miwa T and Okado M: 
Local state probabilities of solvable lattice models: 
An $A^{(1)}_{n}$ family, {\it Nucl. Phys.} 
{\bf B300}[FS22] (1988) 74--108. 
\bibitem{MW}Miwa T and Weston R:
Boundary ABF Models, {\it Nucl. Phys.} {\bf B486 [PM]} 
(1997) 517--545.
\bibitem{BFKZ}Batchelor M T, Fridkin V, Kuniba A 
and Zhou Y.-K: Solutions of the reflection equation 
for face and vertex models associated with 
$A^{(1)}_{n}, B^{(1)}_{n}, C^{(1)}_{n}, D^{(1)}_{n}$ 
and $A^{(2)}_{n}$, {\it Phys. Lett.} {\bf B376} 
(1996) 266--274. 
\bibitem{BFKSZ}Batchelor M T, Fridkin V, Kuniba A, 
Sakai K and Zhou Y.-K: Free energies and critical 
exponents of the $A^{(1)}_1$, $B^{(1)}_n$, 
$C^{(1)}_n$ and $D^{(1)}_n$ face models, 
{\it J. Phys. Soc. Jpn.} {\bf 66} (1997) 913--916. 
\bibitem{H}Hara Y:
Correlation functions of the XYZ model with a boundary,
{\it Nucl. Phys.} {\bf B572 [FS]} (2000) 574--608. 
\bibitem{JKKKM}Jimbo M, Kedem R, Kojima T, Konno H and
Miwa T: XXZ chain with a boundary, 
{\it Nucl. Phys.} {\bf B441 [FS]} (1995) 437--470.
\bibitem{LuP} Lukyanov S and Pugai Ya: 
Multi-point local height probabilities in the 
integrable RSOS model, 
{\it Nucl. Phys.} {\bf B473}[FS] (1996) 631--658. 
\bibitem{LaP}Lashkevich M and Pugai L: 
Free field construction for correlation 
functions of the eight vertex model, 
{\it Nucl. Phys.} {\bf B516} (1998) 623--651. 
\bibitem{AJMP} Asai Y, Jimbo M, Miwa T 
and Pugai Ya: Bosonization of vertex operators 
for the $A^{(1)}_{n-1}$ face model, 
{\it J. Phys. A: Math. Gen.} {\bf 29} (1996) 
6595--6616. 
\bibitem{JKMO}Jimbo M, Kuniba A, Miwa T and Okado M: 
The $A^{(1)}_{n}$ face model, 
{\it Commun. Math. Phys.} {\bf 119} (1988) 543--565. 
\bibitem{MN}Mezincescu L and Nepomechie R I, 
Fusion procedure for open chains, 
{\it J. Phys. A: Math. Gen.} {\bf 25} (1992) 2533--2543.
\bibitem{HSY}Hou B.-Y, Shi K.-J and Yang W.-L, 
The general crossing relation for 
boundary reflection matrix, {\it Commun. Theor. Phys.} 
{\bf 30} (1998) 415--419. 
\bibitem{FF} Feigin B L and Frenkel E V: 
Quantum ${\cal W}$--algebras and elliptic algebras, 
{\it Commun. Math. Phys.} {\bf 178} (1996) 653--678. 
\bibitem{AKOS} Awata H, Kubo H, Odake S and 
Shiraishi J: Quantum ${\cal W}_N$ algebras 
and Macdonald polynomials, 
{\it Commun. Math. Phys.} {\bf 179} (1996) 401--416. 
\end{thebibliography}
\end{document}